\documentclass[10pt,aps,prd,nofootinbib,superscriptaddress,twoside,twocolumn,floatfix,a4paper,eprint,preprintnumbers,showkeys]{revtex4-2}
\usepackage{xcolor}

\usepackage{epsfig,mathtools,bm,color,xcolor,graphicx,subcaption,braket,adjustbox,esint,upgreek,dcolumn,psfrag}

\usepackage{caption}
\usepackage{hyphenat}
\usepackage{amsmath,amssymb,amsfonts}
\usepackage[utf8]{inputenc}
\usepackage[english]{babel}
\usepackage{slashed}
\usepackage{orcidlink}
\usepackage{multirow} 

\usepackage{soul,cancel}

\allowdisplaybreaks 

\usepackage{hyperref}
\hypersetup{colorlinks, linkcolor = [rgb]{0,0,1}, citecolor = [rgb]{0,0.7,0}, urlcolor = [rgb]{0,0,0.8}, pdftitle={}}

\usepackage{cleveref}

\newcommand{\ijs}{\affiliation{Jo{\v z}ef Stefan Institute, Jamova 39, 1000, Ljubljana, Slovenia}}

\newcommand{\ulj}{\affiliation{Faculty of Mathematics and Physics, University of Ljubljana, 1000 Ljubljana, Slovenia}}

\newcommand{\reg}{\affiliation{Institut f{\" u}r Theoretische Physik, Universit{\" a}t Regensburg, 93040 Regensburg, Germany}}

\newcommand{\IMSc}{\affiliation{The Institute of Mathematical Sciences, CIT Campus, Chennai, 600113, India}}
\newcommand{\HBNI}{\affiliation{Homi Bhabha National Institute, Training School Complex, Anushaktinagar, Mumbai 400094, India}}

\newcommand{\msu}{\affiliation{Department of Physics and Astronomy, Michigan State University, East Lansing, 48824, MI, USA}}

\begin{document}

\title{Doubly heavy tetraquarks from lattice QCD: incorporating  diquark-antidiquark operators and the left-hand cut}

\author{Sasa Prelovsek}%\orcidlink{0000-0002-7496-6188}}
\email{sasa.prelovsek@ijs.si}
\ijs \ulj

\author{Emmanuel Ortiz-Pacheco}
\email{ortizpac@msu.edu}
\ijs \msu

\author{Sara Collins}%\orcidlink{0000-0003-0979-7602}}
\email{sara.collins@ur.de}
\reg

\author{Luka Leskovec} %\orcidlink{0000-0002-8926-527X}
\email{luka.leskovec@ijs.si}
\ijs \ulj

\author{M. Padmanath}%\orcidlink{0000-0001-6877-7578}}
\email{padmanath@imsc.res.in}
\IMSc \HBNI

\author{Ivan Vujmilovic}
\email{ivan.vujmilovic@ijs.si}
\ijs \ulj

\begin{abstract}
Lattice studies of the doubly-charm tetraquark $T_{cc}=cc\bar u\bar d$ require the determination of the $DD^*$ scattering amplitude, which most often incorporate only meson-meson interpolators. We additionally incorporate diquark-antidiquark operators and find that these have some impact on certain eigenenergies. This study presents the first extraction of the $DD^*$ scattering amplitude based on the meson-meson as well as  diquark antidiquark interpolators. The effect of the additional operators renders
 slightly smaller values of $p\cot \delta_0$  and a $T_{cc}$ pole slightly closer to the threshold. The scattering amplitude is extracted from eigenenergies by adopting plane-wave and effective-field-theoretic methods, which also incorporate the left-hand cut and address the partial wave mixing. The $T_{cc}$ is found to be a subthreshold resonance with a pole at $m_{T_{cc}}-m_D-m_{D^*}=-5.2^{+0.7}_{-0.8} - i \cdot 6.3^{+2.4}_{-4.8}~$MeV,  employing CLS ensembles with  $m_\pi\simeq 280~$MeV and the distillation method. A more significant effect of diquark-antidiquark operators on eigen-energies is found for larger heavy quark masses relevant for $T_{bb}$. We find that deeply bound $T_{bb}$ does not emerge when employing only meson-meson operators, where each meson is separately momentum projected, while the deeply bound state emerges after adding local diquark antidiquark operators.     \\

\end{abstract}

\maketitle

\section{Introduction}
\label{sec:intro}

The doubly heavy tetraquarks $QQ\bar u\bar d$ with $Q=c,b$ have been the subject of intensive research since the discovery of $T_{cc}=cc\bar u\bar d$ by LHCb collaboration in 2021 \cite{LHCb:2021vvq, LHCb:2021auc}. This exotic hadron is a resonance located less than $1~$MeV below the $D^{*+}D^{0}$ threshold and has a narrow decay width to   $D^0D^0\pi^{+}$. It has isospin $I=0$, while its spin and parity are theoretically expected to be $J^P=1^+$, although these have not yet been experimentally confirmed. The isospin $I=1$ counterpart of this channel has been investigated in the lattice QCD study \cite{Meng:2025}, where a small negative scattering length and no near-threshold poles in the scattering amplitude have been found, consistent with LHCb results.
\vspace{0.2cm}

This paper  investigates $T_{cc}=cc\bar u\bar d$ and explores specific aspects of  $T_{bb}=bb\bar u \bar d$, focusing on the $J^P\!=\!1^+$ and $I\!=\!0$ channel in both cases:

\vspace{0.2cm}

\underline{$T_{cc}$}:  Due to its
 closeness to the threshold, lattice studies need to determine the $DD^*$ scattering amplitude and extract the mass of $T_{cc}$ from the pole location. So far, most lattice determinations of the $DD^*$ scattering amplitude have incorporated only meson-meson bilocal interpolators $D^{(*)}(\vec p_1)D^{*}(\vec p_2)$, where each color-singlet meson is projected to a given momentum \cite{Padmanath:2022cvl,Chen:2022vpo,Lyu:2023xro,Collins:2024sfi,Whyte:2024ihh}. One of the aims of this work is to explore the effect of adding localized diquark-antidiquark operators $[cc]_{\bar 3_c}[\bar u\bar d]_{3_c}$ to the basis. These operators have already been incorporated in  \cite{Cheung:2017tnt,Junnarkar:2018twb,Ortiz-Pacheco:2023ble,Stump:2024lqx} where the scattering amplitude has not been extracted, as well as in our preliminary proceedings \cite{Vujmilovic:2024snz}  where the scattering amplitude was extracted.  

All lattice simulations of $T_{cc}$ are performed at larger-than-physical pion masses, where $m_\pi>m_{D^*} - m_D$ holds and the $D^{*}$ meson is stable. Such kinematics induces a left-hand cut in the partial-wave projected $DD^*$ scattering amplitude, with an associated branch point at real energies immediately below the $DD^*$ threshold, which is a well-known consequence of one-pion exchange in the $u$ channel \cite{Du:2023hlu, Meng:2023bmz}. This invalidates the use of the usual Lüscher's finite-volume formalism \cite{Luscher:1986pf,Briceno:2017max} for extracting the scattering amplitude from lattice eigen-energies along this cut. In this work, this issue is addressed by using an alternative formalism. We adopt an effective potential description of $DD^*$ scattering and solve the Lippmann–Schwinger equation, following Ref. \cite{Meng:2023bmz}: first in finite volume using the plane-wave basis \cite{Meng:2021uhz} in order to fit the parameters of the potential to lattice data, and then finally in infinite volume to find the pole in the scattering amplitude. 

\vspace{0.2cm}

 \underline{$T_{bb}$}: This
 tetraquark has not been experimentally discovered yet, however, a number of reliable theory predictions expect it to be $\approx 100$ MeV below $BB^*$ threshold. Given that it is a bound state well below the threshold, lattice simulations can determine its mass $m\!=\!E_1(P\!=\!0)$ most often based on the ground state energy in finite-volume, e.g. \cite{Francis:2016hui,Junnarkar:2018twb,Leskovec:2019ioa,Alexandrou:2024iwi,Colquhoun:2024jzh,Tripathy:2025vao}. A variety of interpolators have been employed to extract the ground state energy. Our present study shows that bilocal meson-meson operators $B^{(*)}(\vec p_1)B^{*}(\vec p_2)$ alone do not reliably provide the energy in the case of large binding, likely due to small overlap to the localized deeply bound $T_{bb}$. We find that the inclusion of localized diquark-antidiquark operators significantly affects certain energies; in particular, it shifts the ground state eigenenergy downwards. 
The bilocal meson-meson operators and diquark-antidiquark operators have already been employed together previously \cite{Leskovec:2019ioa,Alexandrou:2024iwi}, however, the energies based solely on bilocal meson-meson operators were not provided\footnote{Bilocal operators $O^{MM}$ are called scattering operators in \cite{Leskovec:2019ioa,Alexandrou:2024iwi}.}. 

\vspace{0.2cm}

Our preliminary results incorporating diquark-antidiquark interpolators were presented in \cite{Ortiz-Pacheco:2023ble,Vujmilovic:2024snz}, where the second reference additionally employed the EFT and plane-wave approach to extract the scattering amplitude.  

\section{Lattice Setup  and heavy-light meson masses}

The numerical simulations were performed on two ensembles generated by Coordinated Lattice Simulations (CLS)  ~\cite{Bruno:2014jqa,Bali:2016umi,Bruno:2016plf} with dynamical $u/d$ and $s$ quarks.  They share the same pion mass $m_{\pi} = 280(3) \ \mathrm{MeV}$ and lattice spacing $a$, but have different spatial extents as detailed in Table \ref{tab:lat}.

\begin{table} 
\begin{center}
\begin{tabular}{c|c |c|c|c}
\hline 
\hline
Label & $m_\pi$~[MeV] &   $a$~[fm] & $N_L$ & $N_{cfgs}$ \\\hline
H105 & $280(3)$ &  $0.08636(98)(40)$ & $32$ & 490  \\ 
U101 &  &   &    $24$ & 255\\
  \hline 
\hline
\end{tabular}
\caption{Information on the $N_f=2+1$ CLS lattice ensembles used.}\label{tab:lat} 
\end{center}
\end{table}

\begin{table} 
\begin{center}
\begin{tabular}{c|ccc}
\hline
\hline
$Q=c$ & $m_D=1927(1)~$MeV &~& $m_{D^*}=2049(2)~$MeV\\ 
$Q="b"$ & $m_{B}=4037(3)~$MeV &~& $m_{B^*}=4075(3)~$MeV  \\ \hline \hline
\end{tabular}
\end{center}
\caption{Heavy-light meson masses for two heavy quark masses as determined on the larger volume with $N_L=32$. The employed $"b"$ mass is smaller than the physical $b$-quark mass to avoid too large discretization errors.}
\label{tab:hl-masses} 
\end{table}   

The light and heavy quarks are based on the relativistic nonperturbatively $O(a)$ improved Wilson-Clover action with the clover term set equal to $c_{sw}=1.986246$ for all quarks, and different quark masses realized by different hopping parameters $\kappa$.  Two values of the heavy quark mass are employed: one quark mass is slightly larger than the physical charm quark mass,  and the second one, representing the bottom quark, has a smaller-than-physical $b$-quark mass to avoid too large cutoff effects, which might qualitatively affect the inferences derived here. The employed hopping parameters for light, charm and bottom are $\kappa_{u/d}=0.13697$, $\kappa_c=0.12315$ and $\kappa_{b}=0.0877$, and the relevant pion and heavy-meson masses  are provided in Tables \ref{tab:lat} and  \ref{tab:hl-masses}.  

The results presented are based on fits to the ensemble average, whereas the uncertainties are determined based on the central 68\% distribution of bootstrap samplings. More details on our error analysis are given in Appendix A of \cite{Prelovsek:2020eiw}. 
 
\section{Operator basis}
\label{operator_basis}
Finite-volume energies $E_n$ and overlaps $Z$ of the $QQ\bar u\bar d$ system with $Q=c,b$ are extracted by evaluating all elements of the correlation matrix  
\begin{align}
    C_{ij} (t)& = \langle 0|O_i (t+t_i) \mathcal{O}_j^{\dagger} (t_i)|0 \rangle \nonumber\\
    &= \sum_{n\geq 1} Z_{i}^{n} Z^{n*}_j e^{-E_{n} t}, \quad Z_{i}^{n} \equiv \langle 0| O_i|n\rangle ,
    \label{2ptcorr}
\end{align} 
and solving the generalized eigenvalue problem (GEVP) \cite{Luscher:1990ck,Blossier:2009kd}  
\begin{equation}
\label{gevp}
\begin{split}
 C(t) \mathbf{u}^{(n)}(t&)=\lambda_{n}(t)C(t_0) \mathbf{u}^{(n)}(t)~, \\ &\hspace{-1.05cm} \quad \lambda_n(t)\stackrel{\mathrm{large} ~t}{\rightarrow} A_n e^{-E_n t}~.
\end{split}
\end{equation} 

Below we present the operators employed to create/annihilate the $QQ\bar u \bar d$ system, where for concreteness $Q=c$, while the same set of operators is used also for $Q=b$.  The  meson-meson scattering operators  resemble $DD^*$ and $D^*D^*$, where each color-singlet meson is  projected to a given momentum $\vec{p}_{1,2}$ 
\begin{align} 
    \mathcal{O}^{MM} = &\sum_{\vec{x}_1} e^{i \vec{p}_1 \cdot \vec{x}_1} \bar{u}(x_1) \Gamma_1 c(x_1) \sum_{\vec{x}_2} e^{i \vec{p}_2 \cdot \vec{x}_2} \bar{d}(x_2) \Gamma_2 c(x_2) \nonumber\\
    &- \{ \bar{u} \leftrightarrow \bar{d} \}.
    \label{eq:MM}
\end{align}
The corresponding Dirac and color indices are implicitly contracted, and the total momentum is $\vec P=\vec p_1+\vec p_2$. 
Such bi-local interpolating fields are most commonly used in lattice scattering studies and have also been employed to investigate the $T_{cc}$ system, see e.g. \cite{Cheung:2017tnt,Padmanath:2022cvl, Chen:2022vpo, Collins:2024sfi,Whyte:2024ihh}. 

A tetraquark in a diquark-antidiquark configuration can form a color singlet via  
$ (\mathbf{\bar{3}_c} \otimes \mathbf{3_c})_{\mathbf{1_c}} $ or $ (\mathbf{6_c} \otimes \mathbf{\bar{6}_c})_{\mathbf{1_c}} $.  
The triplet and antitriplet states, $ \mathbf{3_c} $ and $ \mathbf{\bar{3}_c} $, are antisymmetric under color exchange, while the sextet states, $ \mathbf{6_c} $ and $ \mathbf{\bar{6}_c} $, are symmetric.  
Several studies suggest that the dominant contribution to the energy spectrum comes from the $ [cc]_{\bar{3}_c} [\bar{u} \bar{d}]_{3_c} $ configuration \cite{Jaffe:2004ph}.  We employ local diquark-antidiquark operators where all quarks reside at the same position $\vec{x}$ 
\begin{align} 
    \mathcal{O}^{4q}  &= \sum_{\vec{x}} \epsilon^{abc}\epsilon^{ade} \left[ c^{b}_{\alpha} (\vec{x}) \tilde \Gamma_1^{\alpha \beta} c^{c}_{\beta} (\vec{x}) \right] \left[ \bar{u}^{d}_{\delta} \tilde \Gamma_2^{\delta \sigma} \bar{d}^{e}_{\sigma} \right] e^{i \vec{P} \cdot \vec{x}} \nonumber\\
    &\equiv [c \tilde \Gamma_1 c] [\bar u \tilde \Gamma_2 \bar d](\vec P) .
    \label{eq:4q}
\end{align}
Possible effects of local four-quark operators (\ref{eq:4q}) have not been explored extensively in lattice studies at masses closer to that of the charm quark mass. Their effect on eigen-energies in the $T_{cc}$ channel has been found to be insignificant \cite{Cheung:2017tnt,Junnarkar:2018twb} or mild  \cite{Ortiz-Pacheco:2023ble,Vujmilovic:2024snz,Stump:2024lqx}. The present paper represents the first study where the effect of diquark antidiquark operators on the extracted $DD^*$ scattering amplitude is explored.  Preliminary results of this study, which showed the local four-quark operators may have a mild effect, were presented in our proceedings \cite{Vujmilovic:2024snz}.

With the aim to reliably extract finite-volume energies related to $DD^*$ in partial waves $\ell=0,1$, at least up to the lowest inelastic threshold $D^*D^*$, we incorporate the following interpolating fields with total momenta $|\vec P|=0,1\cdot \tfrac{2\pi}{L}$ :
\begin{align}
  \label{ops}
  &\underline{ T_1^+,\  \vec{P}= \vec 0,\  \mathrm{row~z}  \ (J^P\!=\!1^+, \ DD^*_{\ell=0,2})}:  \\
O_1&=O^{D(0)D^*(0)}_{\ell=0} =  \bar q\gamma_{5}c~ (\vec{0})~ \bar  q \gamma_{z}c ~(\vec{0}) ,   \nonumber \\
O_2&=O^{D(0)D^*(0)} _{\ell=0}=  \bar q\gamma_{5}\gamma_{t}c~ (\vec{0})~ \bar  q \gamma_{z}\gamma_{t}c ~(\vec{0}),  \nonumber\\
O_3&=O^{D(1)D^*(-1)}_{\ell=0} =  
\tfrac{1}{\sqrt{6}}   \!\!\!\!\!\!\!\!\!\sum_{\hat{e}_i=\pm \hat{e}_{x,y,z} } \!\!  \!\!\!\! \bar q\gamma_{5}c~ (\hat{e}_i)~ \bar  q \gamma_{z}c ~(-\hat{e}_i),    \nonumber\\
O_4&=O^{D(1)D^*(-1)}_{\ell=2} =\nonumber\\
\tfrac{1}{\sqrt{12}} [& \bar q\gamma_{5}c~ (\hat{e}_x)~ \bar  q \gamma_{z}c ~(-\hat{e}_x) + \bar q\gamma_{5}c~ (-\hat{e}_x)~ \bar  q \gamma_{z}c ~(\hat{e}_x) \nonumber \\
 +& \bar q\gamma_{5}c~ (\hat{e}_y)~ \bar  q\gamma_{z}c ~(-\hat{e}_y) + \bar q\gamma_{5}c~ (-\hat{e}_y)~ \bar  q \gamma_{z}c ~(\hat{e}_y)  \nonumber\\
-2 &\bar q\gamma_{5}c~ (\hat{e}_z)~ \bar  q \gamma_{z}c ~(-\hat{e}_z) -2 \bar q\gamma_{5}c~ (-\hat{e}_z)~ \bar  q \gamma_{z}c ~(\hat{e}_z) ]  ,\nonumber\\ 
O_5&= O^{D^*(0)D^*(0)}_{\ell=0} =  \bar q\gamma_{x}c~ (\vec{0})~ \bar  q \gamma_{y}c ~(\vec{0}), \nonumber\\ 
O_6&= O^{4q}=[c C\gamma_z c] [\bar u C\gamma_5 \bar d](\vec 0). \nonumber\\ 
~&\nonumber\\
&\underline{A_1^-, \vec P=\vec 0 \ (J^P\!=\!0^-, \ DD^*_{\ell=1})}:  \nonumber\\
O_1&=O^{D(1) D^*(-1)}_{\ell=1} =  \nonumber\\
\tfrac{1}{\sqrt{6}} [& \bar q\gamma_{5}c~ (\hat{e}_x)~ \bar  q \gamma_{x}c ~(-\hat{e}_x) - \bar q\gamma_{5}c~ (-\hat{e}_x)~ \bar q \gamma_{x}c ~(\hat{e}_x)  \nonumber\\
 + &\bar q\gamma_{5}c~ (\hat{e}_y)~ \bar  q \gamma_{y}c ~(-\hat{e}_y) - \bar q\gamma_{5}c~ (-\hat{e}_y)~ \bar  q \gamma_{y}c ~(\hat{e}_y) \nonumber\\
  + &\bar q\gamma_{5}c~ (\hat{e}_z)~ \bar  q \gamma_{z}c ~(-\hat{e}_z) - \bar u\gamma_{5}c~ (-\hat{e}_z)~ \bar  q \gamma_{z}c ~(\hat{e}_z)],     \nonumber\\ 
O_2&=O^{D(1) D^*(-1)}_{\ell=1} =O_1(\gamma_{5}\to \gamma_{5}\gamma_t, \gamma_{i}\to \gamma_{i}\gamma_t),\nonumber\\
O_3&=O^{ 4q}= [c C\gamma_t c] [\bar u C\gamma_5 \bar d](\vec 0). \nonumber\\ 
&\nonumber \\ 
&\underline{A_2, \vec P=\tfrac{2\pi}{L} \hat e_z\  \ (J^P\!=\!0^-,1^+,2^-, \ DD^*_{\ell=0,1,2})}:\nonumber\\
 O_1&=O^{D(0) D^*(1)} =  \bar q\gamma_{5}c~ (\vec{0})~ \bar  q \gamma_{z}c ~(\hat{e}_z),  \nonumber   \\ 
 O_2&=O^{D(0) D^*(1)} =  \bar q\gamma_{5}\gamma_{t}c~ (\vec{0})~ \bar  q \gamma_{z}\gamma_{t}c ~(\hat{e}_z),   \nonumber\\ 
 O_3&=O^{D(1) D^*(0)}=  \bar q\gamma_{5}c~ (\hat{e}_z)~ \bar  q \gamma_{z}c ~(\vec{0}),  \nonumber \\ 
 O_4&=O^{D(1) D^*(0)}=   \bar q\gamma_{5}\gamma_{t}c~ (\hat{e}_z)~ \bar  q \gamma_{z}\gamma_{t}c ~(\vec{0}),  \nonumber \\
  O_5&= O^{D^*(1) D^*(0)}= \tfrac{1}{\sqrt{2}} [ \bar q\gamma_{x}c~ (\hat{e}_z)~ \bar  q \gamma_{y}c ~(\vec{0})  \nonumber\\
  &\qquad\qquad \qquad\qquad -  \bar q\gamma_{y}c~ (\hat{e}_z)~ \bar  q \gamma_{x}c ~(\vec{0}) ],  \nonumber\\
  O_6&=O^{ 4q}= [c C\gamma_z c] [\bar u C\gamma_5 \bar d](\hat e_z) . \nonumber
\end{align}
 Light flavors $\bar q\bar q$ in meson-meson operators indicate the isospin 0 combination $\bar q\bar q\to \bar u \bar d- \bar d\bar u$ as in  
(\ref{eq:MM}). 

\section{Correlators with meson-meson and local four-quark operators within distillation}
\label{sec:dist}
We employ the widely used distillation method \cite{HadronSpectrum:2009krc} where all quarks in operators (\ref{ops}) are smeared by applying the  Heaviside Laplacian operator on the point-like quark fields $q_p$ 
{\small
\begin{align}
\label{qs}
&q^{\alpha c}(\vec x,t)\equiv \square_{\vec xc,\vec x^\prime c^\prime} ~q_{p}^{\alpha c^\prime }(\vec x^{\prime},t)=\sum_{k=1}^{N_v} v_{\vec xc}^{(k)}(t)v_{\vec x^\prime c^\prime}^{(k)*}(t)q_{p}^{\alpha c^\prime }(\vec x^{\prime},t)~,\nonumber\\
&\quad N_v^{MM}= \biggl\{ {  60  \  (N_L\!=\!24) \atop 
   90 \  (N_L\!=\!32)  }\ ,\quad  N_v^{4q}= \biggl\{ {  45  \  (N_L\!=\!24) \atop 
   55 \  (N_L\!=\!32)  }.
\end{align}
}  
Our implementation of the Laplacian Heaviside smearing on the quark fields is detailed in \cite{Padmanath:2018tuc,Piemonte:2019cbi}. The quarks fields in the local four quark operators  (\ref{eq:4q})  employ a smaller number of eigenvectors than the meson-meson operators ($N_v^{4q}<N_v^{MM}$) since the 
computational cost is dominated by the matrix element  $\langle O^{4q}|O^{4q\dagger} \rangle$ and rapidly increases with $N_v^{4q}$. Note that smaller number of eigenvectors corresponds to a wider smearing.     

\begin{figure*}[t!]
    \centering
    \begin{subfigure}[t]{0.47\textwidth}
        \centering
        \includegraphics[width=\textwidth]{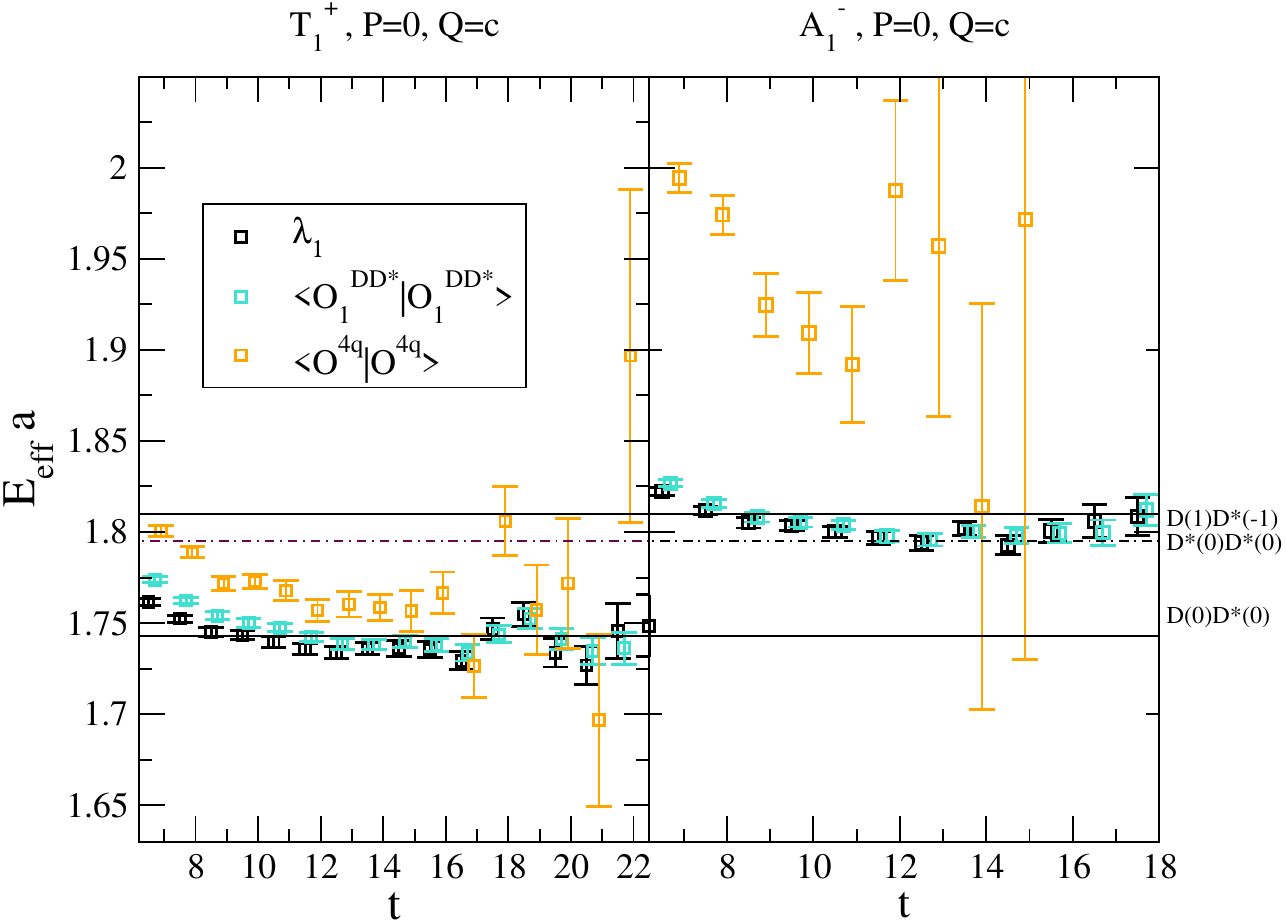}
    \end{subfigure}
    \hspace{0.45cm}
    \begin{subfigure}[t]{0.47\textwidth}
        \centering
        \includegraphics[width=\textwidth]{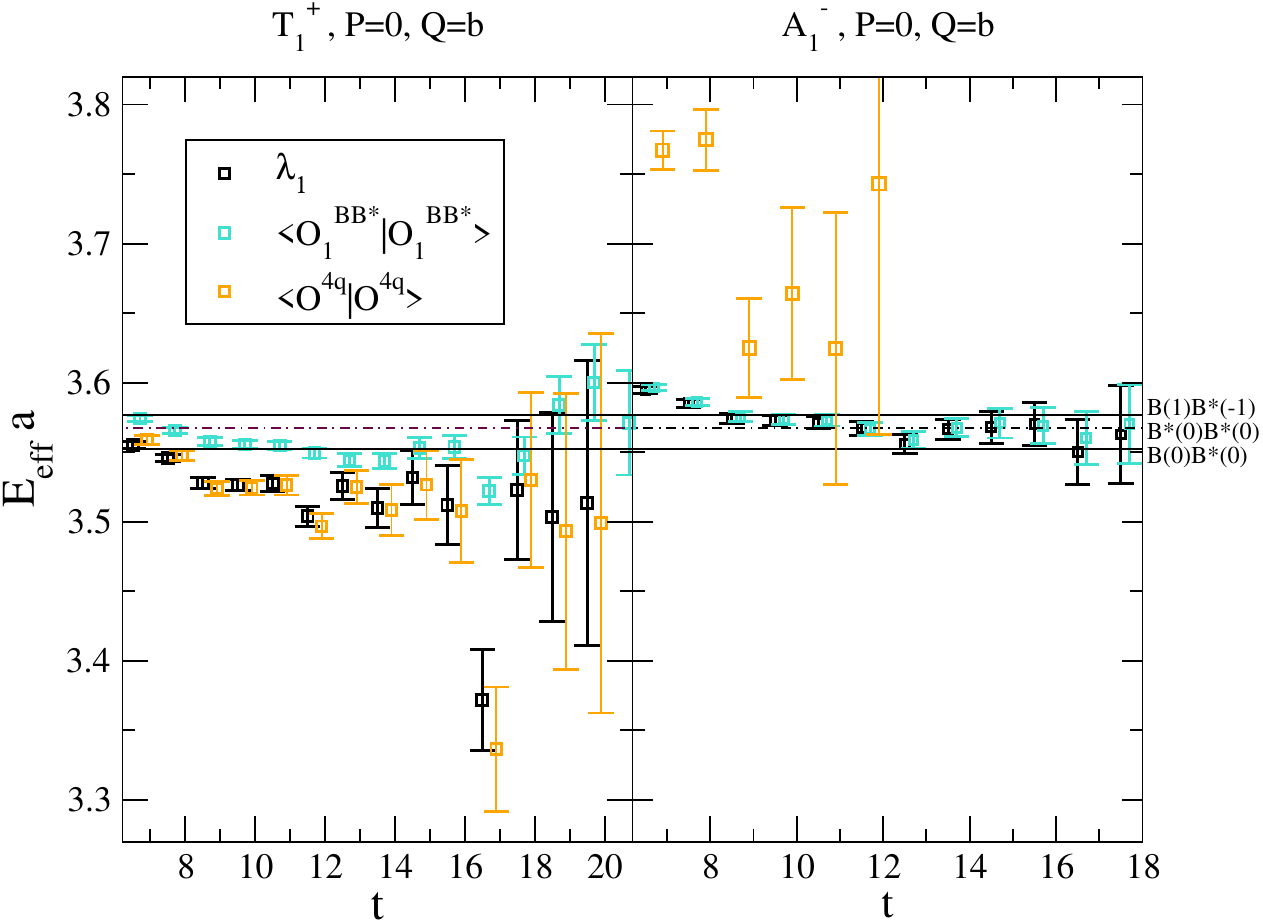}
    \end{subfigure}
    \caption{ The effective energies of the diagonal correlators $\langle O^{4q}| O^{4q}\rangle$ and $\langle O^{DD^*}| O^{DD^*}\rangle$ and  the GEVP ground state eigenvalue $\lambda_1$. Results are shown for two irreducible representations ($T_1^+$, $A_1^-$) and two heavy quark masses ($Q=c,b$) on the smaller volume ensemble~($N_L=24$).   }
   \label{fig:diagonal}
\end{figure*}

 The elements of the correlation matrix (\ref{2ptcorr}) are computed from the following three tensors that were precalculated and stored: quark perambulators $\tau^{kk'}$ that correspond to the propagator from eigenvector $v^{k'}$ to eigenvector $v^{k}$, rank-2 meson matrices $\phi^{jk}$ and the rank-4 tetraquark matrices  $\phi_{4q}^{jklm}$:
\begin{align}
\label{tensors}
\tau^{kk^\prime }_{\alpha \alpha^\prime }(t,t^\prime)&=\sum_{\vec x,c,\vec x^\prime,c^\prime}   v_{\vec xc}^{k*}(t)  ~(D^{-1})^{c,c^\prime}_{\alpha,\alpha^\prime}(t,\vec x;t^\prime,\vec x^\prime)~v_{\vec x^\prime c^\prime}^{k^\prime}(t^\prime ),\nonumber\\ 
&\phi^{jk}(\vec p,t)=\sum_{\vec x,c}v_{\vec xc}^{j*} (t)v_{\vec xc}^{k}(t)e^{i\vec p\cdot \vec x}~,\\
\quad \phi_{4q}^{jklm}(\vec P,&t)=\!\!\!\!\!\!\sum_{\vec x,a,b,c,d,e} \!\!\!\!\!\!\epsilon_{abc}\epsilon_{ade} v_{\vec xb}^{j}(t) v_{\vec xc}^{k}(t)v_{\vec xd}^{l*}(t) v_{\vec xe}^{m*}(t)e^{i\vec P\cdot \vec x}~,\nonumber
\end{align}  
where $j,k,l,m$ represent distillation indices, $a,b,c,d,e$ are color indices and $\alpha$, $\alpha'$ are Dirac indices. 
The correlation matrix elements reduce to contractions of $\tau$, $\phi$ and $\phi_{4q}$ tensors (\ref{tensors})  over their respective indices. The sums over the distillation indices running from $1$ to $N_v$ are the most expensive step in the computation. This particularly increases the numerical cost of the calculation of the correlator that involves the tensor $\phi_{4q}$ since it is of rank 4 in the distillation space, or more precisely:
\begin{align}
&\langle O^{4q}(t)|O^{4q\dagger}(t') \rangle=- ~\phi_{4q}^{jklm}(t,\vec P) ~\Gamma_1^{\alpha\beta}~\Gamma_2^{\gamma\delta}~ \nonumber\\
\ \ &\cdot ~\bigl[ ~\tau^{jj'}_{\alpha \alpha '}(t,t')\tau^{kk'}_{\beta \beta '}(t,t') \tau^{ll'}_{\gamma \gamma '}(t,t')\tau^{mm'}_{\delta \delta'}(t,t')\nonumber\\
\ \ &\Gamma_{1'}^{\alpha'\beta'}~\Gamma_{2'}^{\gamma'\delta'}~   \phi_{4q}^{j'k'l'm' *}(t',\vec P) 
 -\{j\leftrightarrow k, \alpha \leftrightarrow \beta \}\bigr],
 \label{C4q}
\end{align}
where the replacement of indices in the third line applies only within the square parenthesis. 

The effective energies of diagonal correlators $\langle O^{4q}| O^{4q \dagger}\rangle$  (\ref{C4q}), along with $\langle O^{DD^*}| O^{DD^* \dagger}\rangle$ and the GEVP ground state effective energies
 are compared in Figure \ref{fig:diagonal}.  The correlators based on local four-quark operators show larger errors than those based on meson-meson operators.  The local operator $O^{4q}$  in the most relevant irreducible representation $T_1^+$ leads to a distinctly lower effective mass than that of the bi-local meson-meson operator $O_1$ in the case of $Q=b$. This is likely related to the important contribution of the diquark-antidiquark Fock component in $T_{bb}$ and the poor overlap of meson-meson scattering operators. The local operator $O^{4q}$  in irrep $A_1^-$ leads to a higher effective mass than the meson-meson operator $O_1$; this is not surprising since the pseudoscalar diquark $[QC\gamma_tQ]$ is not one of the lower-lying diquarks according to Jaffe's classification in Table III of \cite{Jaffe:2004ph}. 

We note that the correlators (\ref{C4q}) involving the local four-quark operators with $N_v^{4q}\simeq 50$ turned out to be at least an order of magnitude more costly in our implementation than the correlators employing just meson-meson operators with $N_v^{MM}\simeq 100$.  A recent proceedings \cite{Stump:2024lqx} proposes a different method that is also based on the precalculated perambulators $\tau$, but avoids tensors of rank-4 in the distillation space \cite{Stump:2024lqx}.
 
\section{Finite-volume energies, overlaps, and impact of local four-quark operators}

This section presents the energies $E_{n}$ of the finite-volume eigenstates and their overlaps to operators $Z_i^n=\langle O_i|n\rangle$, 
 and details the investigation of the impact of local-four quark operators. All results are obtained
employing the GEVP (\ref{gevp}) with $t_0\!=\!4$, using correlators which are averaged over 5 or 8 source time-slices $t_i$ (\ref{2ptcorr}), over three polarizations (for $T_1^+$) or three total momenta (for $|\vec P|=2\pi/L$).  

\begin{figure}[h!]
    \centering
    \includegraphics[width=\columnwidth]{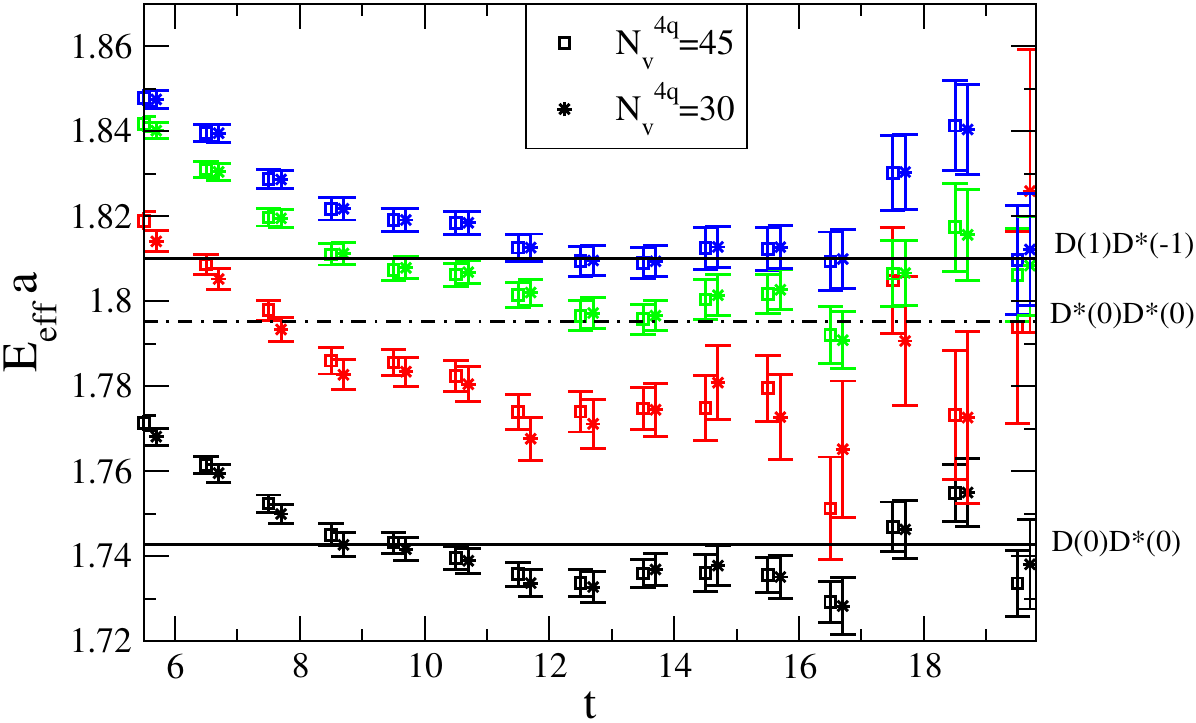}
    \caption{Effective energies of eigenstates for $\vec{P} = \vec{0}$, irrep $T_1^+$ on ensemble U101 ($N_L=24$) employing all six operators. Squares correspond to smearing with  $N_v^{4q}=45$ eigenvectors in $O^{4q}$, while stars correspond to $N_v^{4q}= 30$. In both cases, $N_v^{MM}=60$ is employed. Note that all other results and figures employ the $N_v$ given in  (\ref{qs}).}
   \label{fig:compare-nvs}
\end{figure}

A comparison of the results for two different smearing widths for the quarks in operators $O^{4q}$ in Figure~\ref{fig:compare-nvs} indicates that errors on certain effective energies decrease slightly when going from a larger width ($N_v^{4q}\!=\!30$) to a smaller width ($N_v^{4q}\!=\!45$).  This affects the errors on eigenstates that couple to the local four-quark operators.  

\begin{figure*}[t!]
    \centering
    \begin{subfigure}[h!]{0.7\textwidth}
        \centering
        \includegraphics[width=\textwidth]{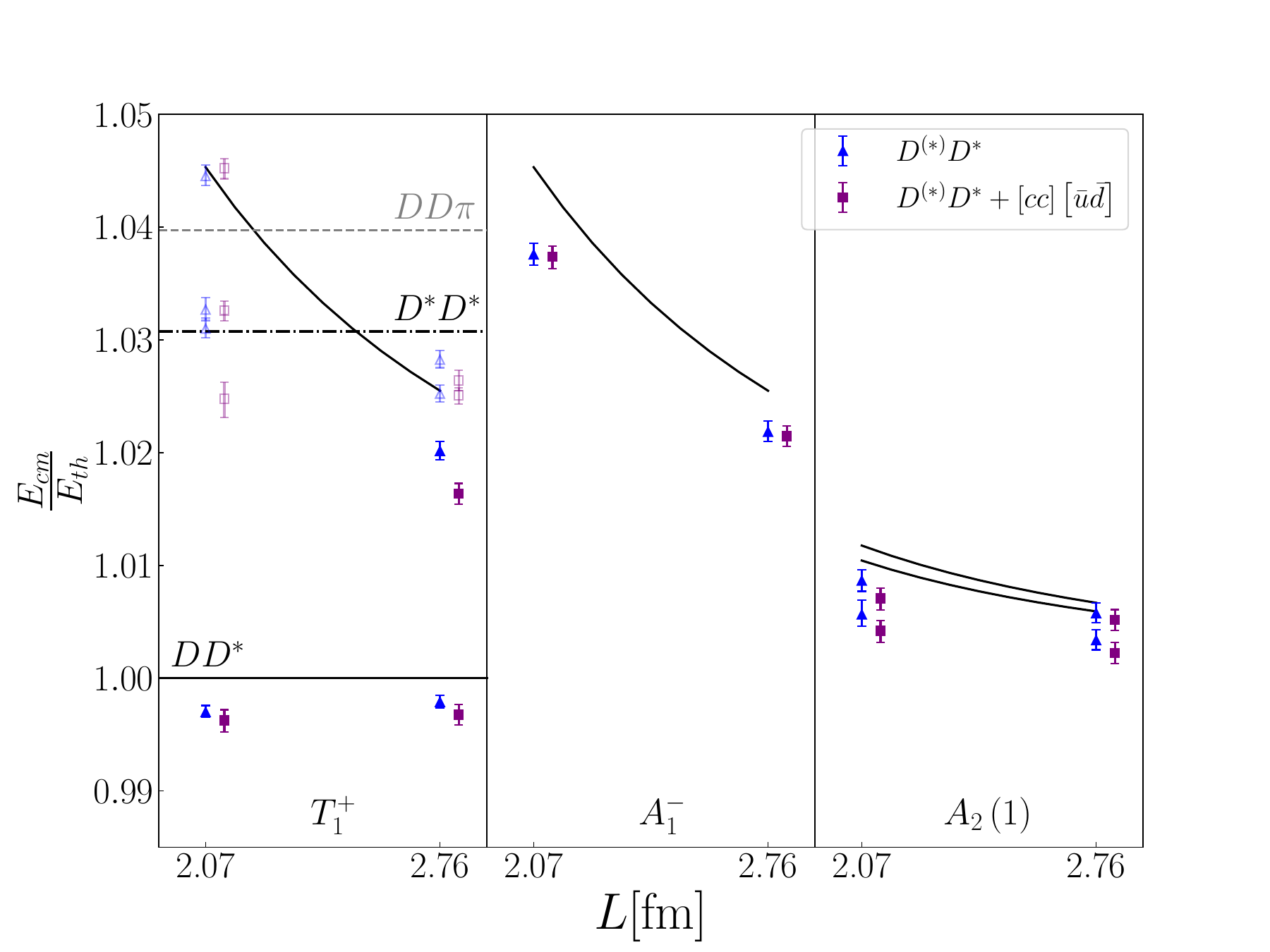}
    \end{subfigure}
    \begin{subfigure}[h!]{0.9\textwidth}
        \centering
        \includegraphics[width=\textwidth]{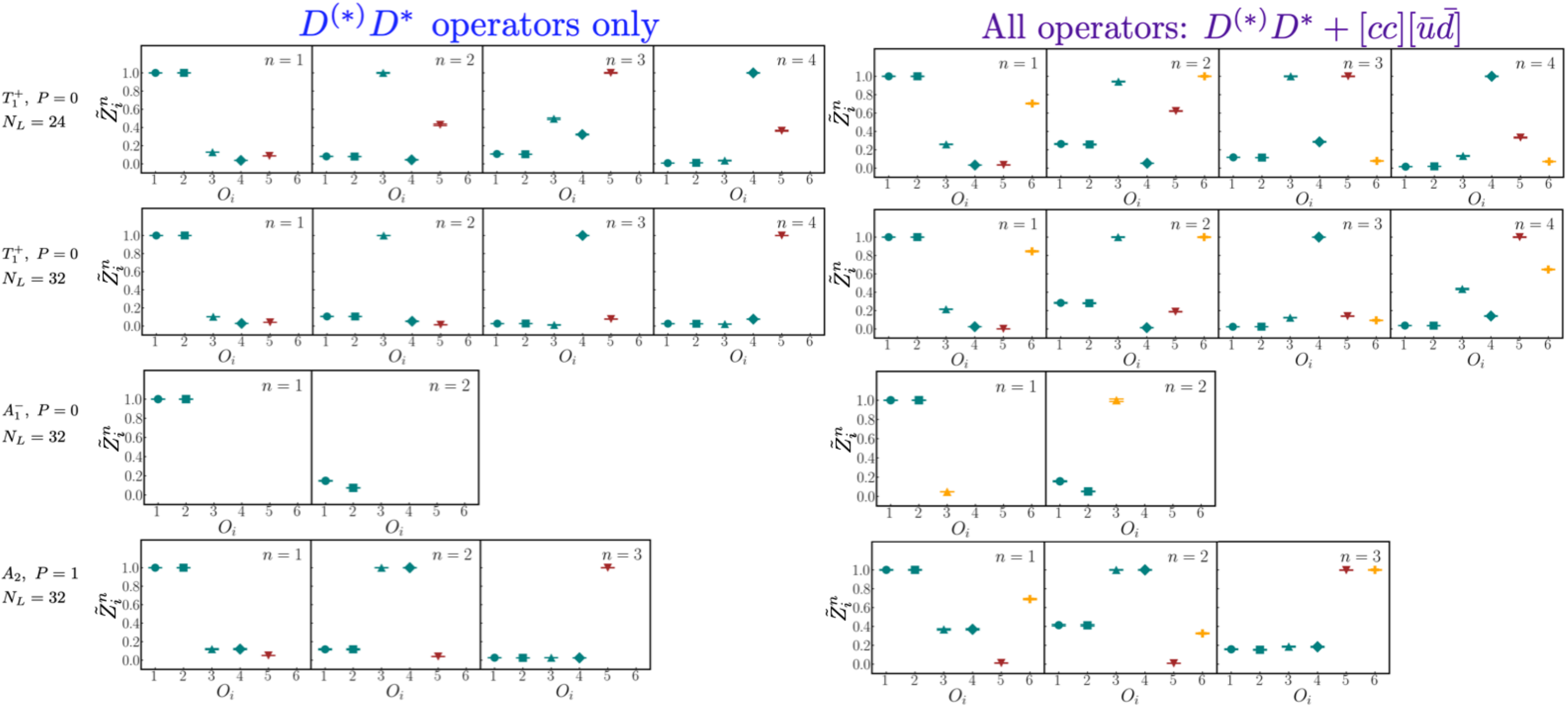}
    \end{subfigure}
    \caption{(a) Finite-volume energy spectrum for $Q=c$, shown in the form of the ratio $E_{cm}/E_{th}$ with $E_{th}=m_D+m_{D^*}$: results including all interpolators (violet), results including only meson-meson interpolators and excluding local four quark operators (blue) and the non-interacting energies (lines) are displayed. 
    The filled symbols indicate energy levels employed in the scattering analysis. 
    (b) The normalized overlaps of eigenstates to employed operators: here $Z_i^n=\langle n|O_i\rangle$, while   $\tilde Z_i^n\equiv Z_i^n/\mathrm{max}_{n'} Z_i^{n'}$     is normalized by the overlap of the operator $O_i$ to the state that has largest overlap to this operator among all eigenstates. Therefore  $\tilde Z$  is independent of the normalization of the operator.}
    \label{fig:en_Z_c}
\end{figure*}

In the following, we present lattice energies as  $E_n=\Delta E_n^{lat}+E_{con}^{ni}$  which will represent an input to the scattering analysis. Here, the energy shifts $\Delta E_n^{lat}$ and the continuum non-interacting energies $E_{con}^{ni}$  are  
\begin{align}
\label{delE}
\Delta E_n^{lat}&=E_n^{lat}-E^{lat}_{D^{(*)}(\vec p_1)}-E^{lat}_{D^{*}(\vec p_2)}\nonumber\\  
E_{con}^{ni}&=(m_{D^{(*)}}^2+\vec p_1^{\,2})^{1/2}+(m_{D^{*}}^{\,2}+\vec{p_2}^{\,2})^{1/2}~.
\end{align}
The combination $\Delta E_n^{lat}+E_{con}^{ni}$  mitigates small deviations of single-hadron energies from their continuum values and ensures that the scattering amplitude is zero if $\Delta E_n^{lat}$ is zero. 

Below we discuss the spectrum and overlaps separately for the two heavy-quark masses employed, as certain findings are quite different: 
\begin{itemize}
\item 
$\mathbf{Q=c}$: The spectrum for the charm sector in Figure \ref{fig:en_Z_c}a compares eigen-energies obtained including meson-meson and diquark-antidiquark operators (violet), eigen-energies obtained including only meson-meson operators  (blue) and non-interacting energies $E^{ni}=E_{D^{(*)}}(\vec p_1)+E_{D^*}(\vec p_2)$ (lines). The energies remain roughly unaffected by the inclusion of diquark-antidiquark operators, i.e. the energies employing basis $D^{(*)}D^*$ and $D^{(*)}D^*+[cc][\bar u\bar d]$  are consistent within the 1$\sigma$   statistical uncertainties. The exception is the second eigenstate in irreducible representation $T_1^+$, whose energy is decreased by a few $\sigma$ with the inclusion of diquark antidiquark operators, as confirmed also in Figure  \ref{fig:dE_c}  that scrutinizes various fit-ranges for this level. The normalized overlaps of eigenstates to employed operators are presented in  \ref{fig:en_Z_c}b. The diquark-antidiquark operator couples to several eigenstates, which is expected as it has the same quantum numbers and is Fierz-related to other interpolators used \cite{Nieves:2003in,Padmanath:2015era}. The pattern of overlaps  $\langle n|O^{MM}\rangle$ remains mostly unaffected with the inclusion of diquark-antidiquark operators. In particular, each level is dominantly coupled to only one of the $DD^*$ or $D^*D^*$ operators (in addition to being possibly coupled also to the diquark-antidiquark operator), which is advantageous for the one-channel $DD^*$ scattering analysis performed below\footnote{The exception to the last two sentences is level $n=3$ in $T_1^+$ on the $N_L=24$ ensemble, which lies very near the $D^*D^*$ threshold and is therefore not included in the scattering analysis below. }. 

\begin{figure}[t!]
    \centering
    \includegraphics[width=1.0\columnwidth]{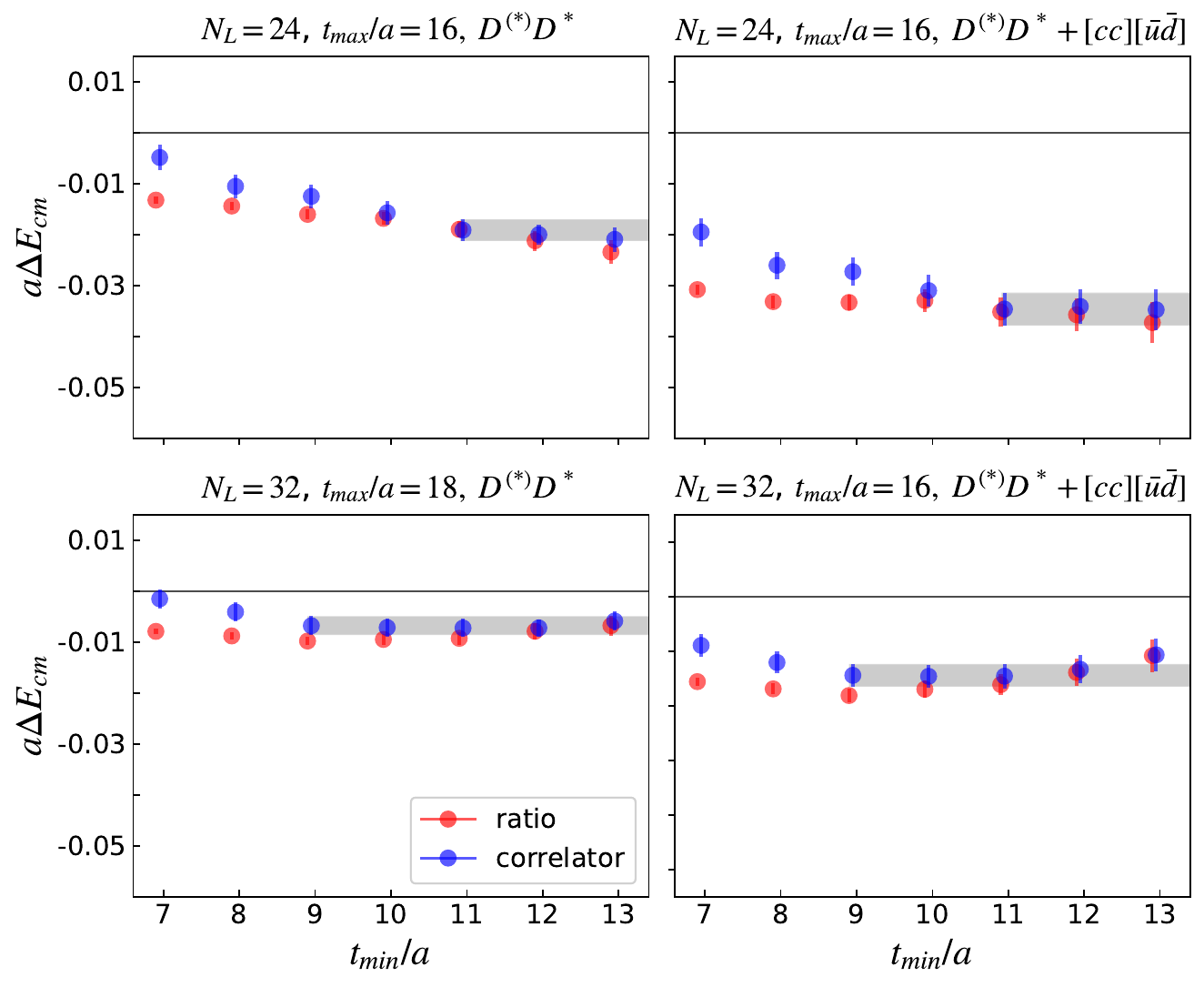}
    \caption{Energy shifts (\ref{delE}) of the first excited eigen-energy in the $T_1^+$ irrep in the charm sector that is affected by the inclusion of $O^{4q}$: $\Delta E=E_{2}-E_{D(1)}-E_{D^*(1)}$.    Results using only meson-meson operators (left)  and using all interpolators (right) are shown. The shifts are displayed as a function of the $t_{min}$ utilized in the one-exponential fit. The gray bands indicate the chosen fit estimates employed in the scattering analysis. They are evaluated using energy estimates from separate fits to each of the correlators involved. The label ``ratio" refers to energy splittings extracted from single fits to the ratio of the interacting eigenvalue correlators to the product of single meson correlators, whereas the label ``correlator" refers to the same quantity extracted from separate fits to each of the correlators involved.  Other levels are not significantly affected by the inclusion of  $O^{4q}$ for $Q=c$. }
    \label{fig:dE_c}
\end{figure}

\item $\mathbf{Q="b"}$: The $BB^*$ and $B^*B^*$ thresholds lie much closer together due to the hyperfine splitting decreasing with increasing heavy quark mass. 
The influence of local four-quark operators is striking for this heavy quark mass.   This is evidenced by the spectrum in Figure \ref{fig:en_Z_b}a, where the pattern of eigen-energies is affected when including the local $[bb][\bar u\bar d]$ operator (violet)  in addition to the bi-local operators $B^{(*)}(\vec p_1)B^*(\vec p_2)$ (blue). The most prominent effect with the inclusion of local four-quark operators in the basis is the observation of a new distinct ground state in the $T_1^+$ and $A_2$ irreps that were inaccessible with purely bilocal meson-meson interpolators. 

\begin{figure*}[t!]
    \centering
    \begin{subfigure}[h!]{0.7\textwidth}
        \centering
        \includegraphics[width=\textwidth]{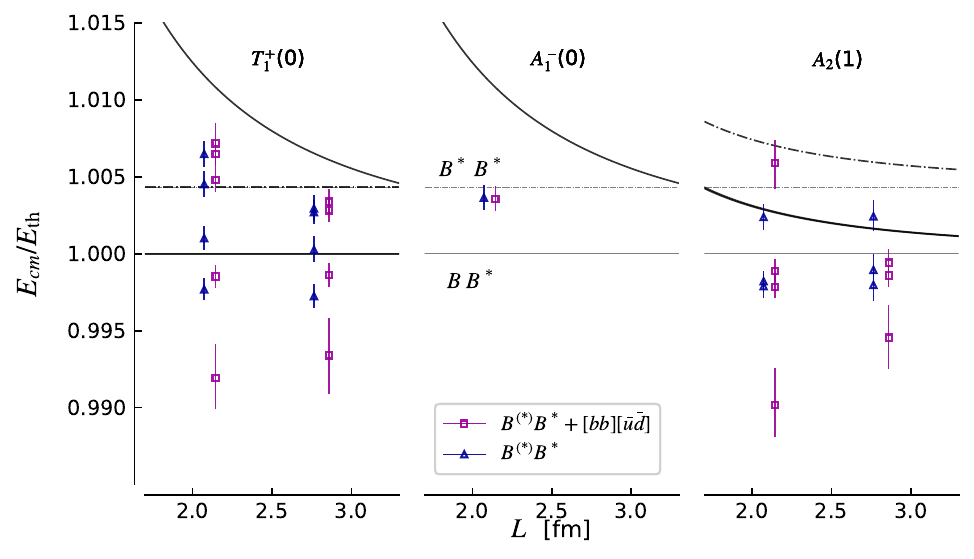}
    \end{subfigure}
    \begin{subfigure}[h!]{0.9\textwidth}
        \centering
        \includegraphics[width=\textwidth]{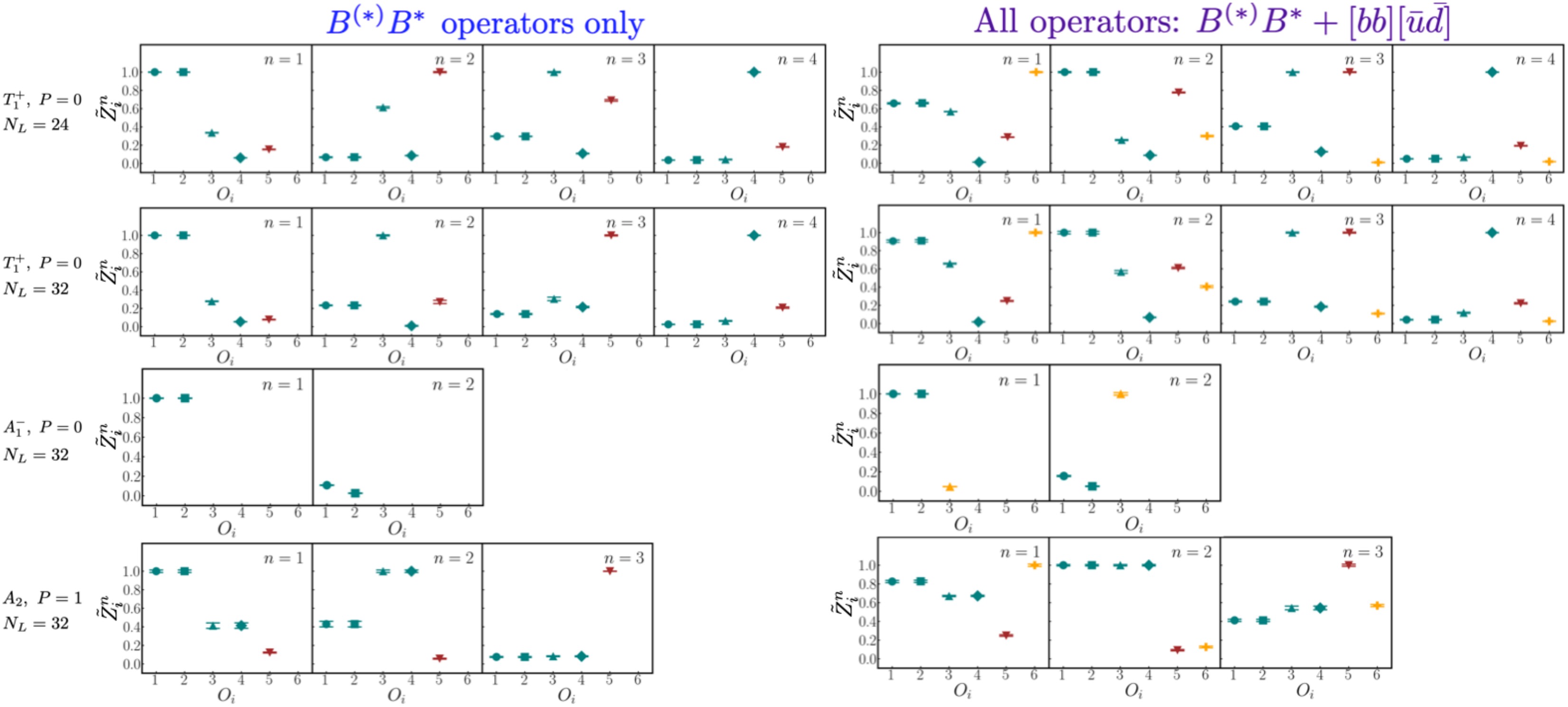}
    \end{subfigure}
    \caption{(a) Finite-volume energy spectrum $E_{cm}/E_{th}$ with $E_{th}=m_B+m_{B^*}$ for $Q=b$: results including all interpolators (violet), results including only meson-meson interpolators and excluding local four quark operators (blue) and non-interacting energies (lines).  The irrep $A_1^-$ has not been simulated on a larger volume to reduce the cost of simulation.    (b) Corresponding normalized overlaps  $\tilde Z_i^n\equiv Z_i^n/\mathrm{max}_{n'} Z_i^{n'}$.
    \label{fig:en_Z_b}}
\end{figure*}

The large statistically significant difference between the ground state energies using a basis with or without the local four-quark operators is evident from the $t_{min}$ dependence of energy splittings presented in Figure \ref{fig:dE_b}. The ground state from the basis omitting local four-quark operators and the first excited state from the basis including them have nearly consistent energies.  This corroborates the idea that the ground state observed using the enlarged basis represents a new distinct level.

\begin{figure}[h!]
    \centering
    \includegraphics[width=1.0\columnwidth]{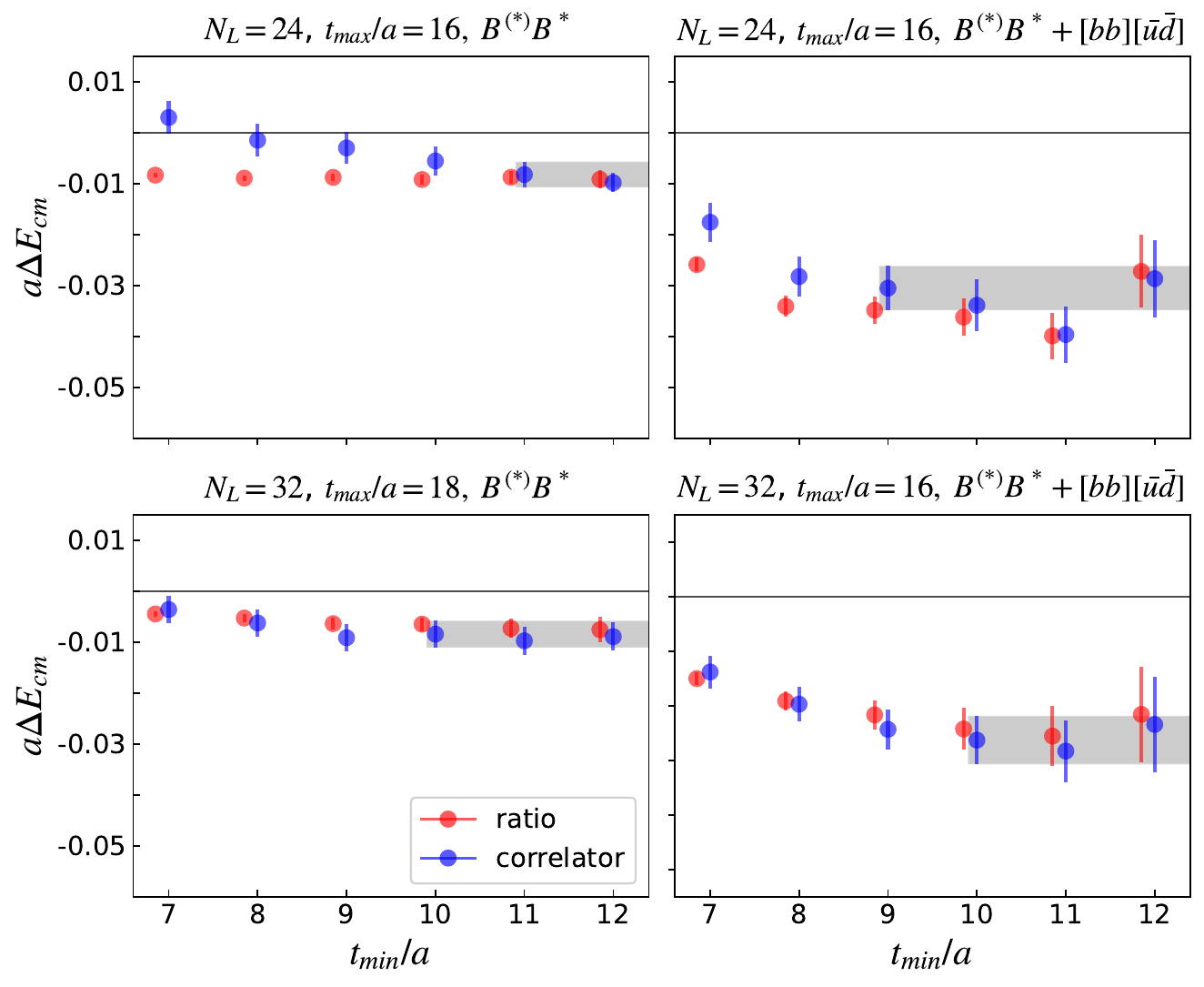}
    \caption{ Energy shift (\ref{delE}) for one of the levels that are significantly affected with the inclusion of $O^{4q}$ in the "bottom" sector: $\Delta E=E_1-m_B-m_{B^*}$  for the ground state in irrep $T_1^+$ as a function of $t_{min}$ in the one-exponential fit. A number of other levels are also significantly affected by the inclusion of  $O^{4q}$ for $Q=b$, as evidenced in Fig. \ref{fig:en_Z_b}. The definitions of red, blue, and gray shifts are detailed in the caption of Figure \ref{fig:dE_c}.   }
    \label{fig:dE_b}
\end{figure}

The operator-state overlaps in Figure \ref{fig:en_Z_b}b also show prominent effects with the inclusion of the diquark-antidiquark operators, supporting the statement above. The ground states in irreps $T_1^+$ and $A_2$ are dominantly coupled to diquark-antidiquark operators, and show no characteristic resemblance to the pattern of overlaps for any levels determined using purely bilocal meson-meson interpolators (see Figure \ref{fig:en_Z_b}b). This is in line with the expectation that the local diquark-antidiquark Fock component plays a dominant role in $T_{bb}$ according to many lattice and phenomenological studies, e.g. \cite{Bicudo:2012qt,Bicudo:2015kna,Francis:2016hui,Junnarkar:2018twb,Leskovec:2019ioa,Alexandrou:2024iwi,Colquhoun:2024jzh,Tripathy:2025vao,Karliner:2017qjm,Janc:2004qn}. The overlap factors in the first excited state in the $T_1^+$ irrep from the enlarged basis can be approximately seen to reflect the patterns for the ground state using purely meson-meson interpolators. Such a comparison of overlaps for higher levels and the levels in the $A_2$ irrep is more complicated as the inclusion of local four-quark operators leads to eigenstates with comparable couplings to both meson-meson operators of type $BB^{*}$ and $B^{*}B^{*}$.

The observation of this lower level with the enlarged basis indicates that bilocal meson-meson interpolators fail to access the ground state in the $T_{bb}$ sector within moderate physical time separations.  
With the inclusion of the local diquark-antidiquark operator, we obtain a $T_{bb}$ binding energy of approximately $60~\pm 10$ MeV for the lighter-than-physical $"b"$ quark we investigate (see Table \ref{tab:hl-masses}). The deep binding of $T_{bb}$ was reported by previous lattice studies \cite{Leskovec:2019ioa, Alexandrou:2024iwi} that used local diquark-antidiquark operators alongside local and bilocal meson-meson interpolators in their analysis. These two studies employed nonrelativistic QCD for bottom quarks, while our present study confirms this finding with relativistic bottom quarks. Several other studies also have reported similar deep binding using only local $BB^*=(b\bar d)(b\bar u)$ operators together with local diquark-antidiquark interpolators \cite{Francis:2016hui,Junnarkar:2018twb,Mohanta:2020eed,Hudspith:2023loy,Colquhoun:2024jzh}\footnote{Lattice investigations following static potential-based and HALQCD-potential based approaches also predict similar binding energies in $T_{bb}$ system \cite{Bicudo:2012qt,Bicudo:2015kna,Aoki:2023nzp}}. The only previous lattice simulation of $T_{bb}$ that evaluates all correlators between bi-local $B(0)B^*(0)$ and local $[bb][\bar u\bar d]$  did not present the result based on bilocal $B(0)B^*(0)$ operators alone \cite{Alexandrou:2024iwi}.  In addition to the heavy-quark treatment mentioned above, there are further differences between  Ref. \cite{Alexandrou:2024iwi} and our study:  the former employed stochastic noise propagators for evaluating Wick contractions and we employ distillation method; the former is based on the staggered action for light quarks and we use Wilson-clover action.  
  
 The finite-volume energies in the bottom sector presented in Fig. \ref{fig:en_Z_b}a have significant statistical errors and are also expected to have significant heavy-quark discretization errors. They are not, therefore, reliable enough to be utilized to extract the scattering amplitudes. In addition, the closeness of the $BB^*$ and $B^*B^*$ thresholds calls for the extraction of the coupled-channel scattering matrix for both channels,  which is beyond the scope of the present study. 
\end{itemize}

\section{$\mathbf{DD^*}$ scattering amplitude from effective field theory and plane-wave approach}

\begin{figure}[t!] 
\begin{center}
        \includegraphics[width=0.5\textwidth]{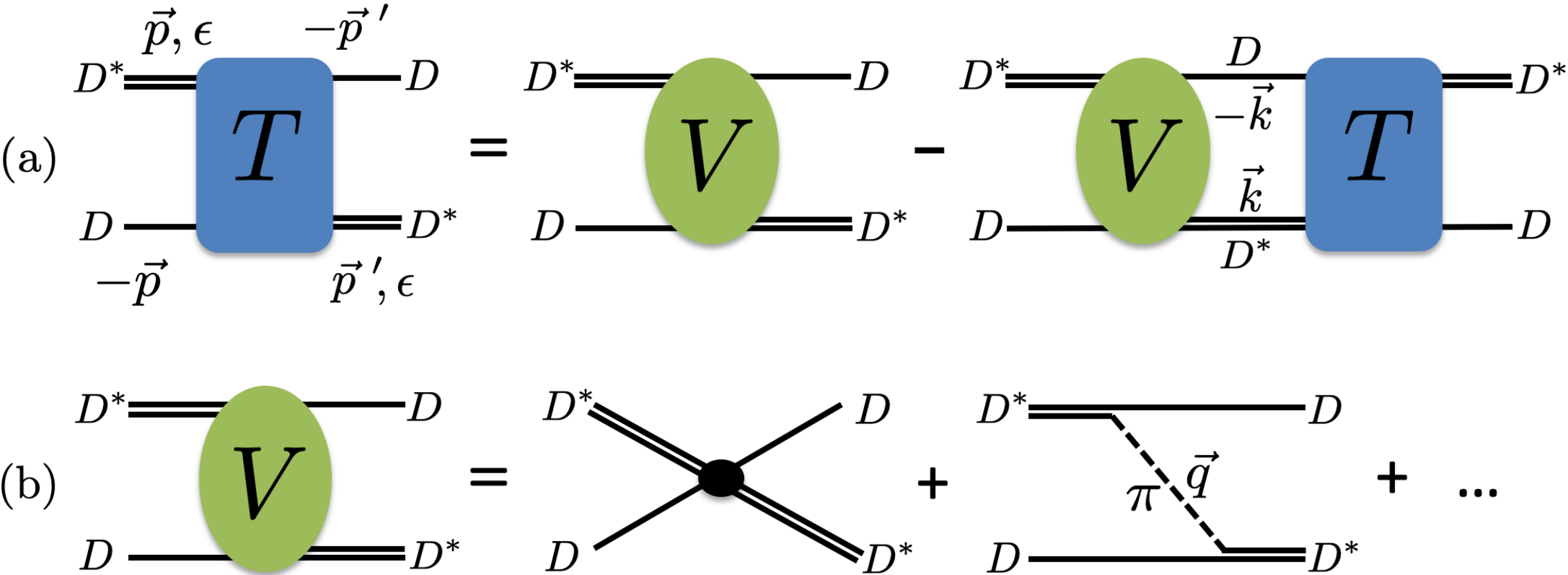}  
    \caption{(a) Lippmann-Schwinger relation between scattering amplitude $T$ and potential $V$, where momenta refer to the center-of-momentum frame.
    (b) Effective potential $V$ defined in (\ref{V})\footnote{Note that signs of the momenta differ with respect to \cite{Collins:2024sfi}.}. }
    \label{fig:effpot}
    \end{center}
\end{figure}

 We aim to determine the $DD^*$ scattering amplitude $T_l$ for the lowest partial waves $l=0,1$. These amplitudes  can be expressed in terms of the scattering phase shift $\delta_l$ as \cite{Du:2023hlu}
\begin{equation}
\label{amplT}
-\frac{2\pi}{m_r}T_l^{-1}=p\cot\delta_l-ip~,
\end{equation}
where $m_r$ is the reduced mass of the $DD^*$ system.  The main obstacle to the applicability of Lüscher's formalism for extracting scattering amplitude is the existence of a left-hand cut.  This results from the one-pion exchange (OPE), illustrated on the right-hand-side of   Figure \ref{fig:effpot}b, when the pion comes on-shell \cite{Du:2023hlu}. For this reason, we will employ an effective field theory approach, where the $DD^*$ potential $V$ is represented by the sum of the one-pion exchange and the local $DD^*$ interaction. The unknown local interaction will be parametrized using several free low-energy constants. These constants are determined by fitting the lattice spectrum to the energies of Hamiltonian  $H=\frac{p^2}{2m_r}+V$ in the finite volume and in the plane wave basis, introduced in \cite{Meng:2021uhz}.  Once the parameters of the potential are known, the scattering amplitude $T$ is determined from the Lippmann-Schwinger equation illustrated in Fig. \ref{fig:effpot}a in infinite volume. 
The kinematics of $D$ and $D^*$ mesons are treated in the non-relativistic limit as in Section 5 of \cite{Meng:2021uhz}  since we consider the scattering amplitude at energies in the close vicinity of the $DD^*$ threshold. 

\subsection{The Lippmann-Schwinger equation}

The scattering amplitude $T$ and $DD^*$ potential $V$ are related via the Lippmann-Schwinger equation (LSE) illustrated in Figure \ref{fig:effpot}a for the center-of-momentum (CMF) frame\footnote{The sign of $G$ differs from \cite{Meng:2021uhz,Vujmilovic:2024snz} and is consistent with \cite{Du:2023hlu,Collins:2024sfi}. }
\begin{align}
 \label{eq:LSE}
  &{T} = {V} -{V} {\mathcal{G}} {T},\\
T( \vec p,\vec p^{\,\prime} ;E)=V( \vec p,\vec p^{\,\prime}&)-\int \!\!\frac{d\vec k}{(2\pi)^3} V( \vec p,\vec k) G( \vec k;E)T( \vec k,\vec p^{\,\prime};E). \nonumber
\end{align} 
The second relation corresponds to the infinite volume, while the first relation can be conveniently expressed in terms of matrices for a finite basis in the finite volume. Our main aim is to extract the infinite-volume on-shell scattering amplitude $T(|\vec p|\!=\!|\vec p^{\,\prime}|\!=\!p;E\!\simeq\! E_{th}\!+\!p^2/2m_r)$, while the off-shell amplitude inherently appears within the integrand.   Here $G$ is the propagator which reduces to the Green's function of the Schrödinger  equation in the nonrelativistic limit (see footnote  quoted before equation (\ref{eq:LSE}))
\begin{align}
    G(p^0, \vec{p}) = \frac{1}{  \frac{\vec{p}^{\,2}}{2 m_r} - p^0+ i \epsilon},
    \label{eq:prop}
\end{align}
and is placed in the plane-wave basis through
\begin{align}
     \mathcal{G}=\frac{\Delta \vec k}{(2\pi)^3}G=\frac{1}{L^3}G.
    \label{eq:prop2}
\end{align}

Poles of the scattering amplitude $T$  (\ref{eq:LSE}) 
\begin{equation}
\label{eq:deteq}
T = \left( {\mathcal{G}}^{-1} + {V} \right)^{-1}  {\mathcal{G}}^{-1}{V}
\end{equation}
 are determined from  
 \begin{equation}
\mathrm{det}\left( {\mathcal{G}}^{-1} + {V} \right) = 0,
\end{equation}
which in turn leads to the familiar Hamiltonian equation when the propagator G defined in \eqref{eq:prop2} is inserted into the determinant equation
\begin{align}
    \mathrm{det} \left( {H} - p^0 {I} \right) = 0,\qquad {H} = \frac{p^2}{2 m_r}I + \frac{1}{L^3}{V}.
    \label{eq:hameq}
\end{align}
Note that the units of the potential $V$ and scattering amplitude $T$ are $1/$GeV$^2$, which renders the correct units (GeV) for the Hamiltonian. The same relation also holds in finite volume where it is projected to appropriate irreducible representations $\Gamma$ of the octahedral group $O_h$ or one of its little groups 
\begin{align}
    \mathrm{det} \left( H^{\Gamma} - p^{0, \Gamma} I \right) = 0 .
    \label{eq:hameqirr}
\end{align}
This equation is fulfilled precisely when $p^{0, \Gamma}$  equals one of the finite-volume energies  $E_n^{cm}$ on the lattice.  This relation is used to extract the free parameters of the potential by fitting to lattice eigen-energies. This approach has been introduced for the two-nucleon case in Section 5 of  \cite{Meng:2021uhz}. 

\subsection{The effective potential}
\label{subsec:effpot}
\indent The application of the Lippmann-Schwinger equation necessitates the use of an effective potential when parametrizing the interaction in the $DD^{*}$ system. This will be composed of a one-pion exchange and a contact $DD^*$ interaction 
\begin{equation}
V=V_\pi+V_{CT}~.
\end{equation}

 The one-pion exchange is incorporated via the effective  Lagrangian \cite{Fleming:2007rp} 
\begin{align}
\label{eq:lagr}
    \mathcal{L}_{int} &= \frac{g}{2 f_{\pi}} \left( D^{* \dagger} \cdot \nabla \pi^{a} \tau^{a} D + \mathrm{h. \ c.} \right), \\  
    \ &\pi^{a} \tau^{a} = \begin{pmatrix}
\pi^{0} & \sqrt{2} \pi^{+} \\
\sqrt{2} \pi^{-} & - \pi^{0}
\end{pmatrix}. \nonumber
\end{align}
While the physical values of the low-energy constants from (\ref{eq:lagr}) are $g^{ph}\simeq 0.57$ and $f_{\pi}^{ph} = 92.2 \ \mathrm{MeV}$, we take their values at $m_\pi\simeq 280~$MeV:  $g = 0.645$ is based on the lattice simulation  \cite{Becirevic:2012pf} and $f_\pi=105~$MeV is based on chiral perturbation theory (see Appendix A.3 in \cite{Du:2023hlu}). The potential between $DD^{*}$ mesons is then given as 
\begin{align}
\label{eq:pot1}
    V_{\pi} ( \vec p,\vec p^{\,\prime}) = 3 \left( \frac{g}{2 f_{\pi}} \right)^{2} \frac{\left( \vec{\epsilon} \cdot \vec{q} \right) \left( \vec{\epsilon}^{\,\prime *} \cdot \vec{q} \right)}{q^2 - m_{\pi}^2} ,
\end{align}
with momentum transfer $\vec q=\vec p+\vec{p}^{\,\prime}$. The s-wave projection of the  potential, derived in \cite{Collins:2024sfi}, features a left-hand cut beginning approximately at $p_{lhc}^{2} \approx - \mu_\pi^2/4\simeq -10^{-3}  E_{th}^2 $ with  $\mu_\pi^2\equiv m_\pi^2-(m_{D^*}-m_D)^2>0$ for our $DD^*$ system. This is the energy below which the exchanged pion can come on shell, and the cut extends to $-\infty$ along the real energy. The central part of the potential $V_\pi$ in (\ref{eq:pot1}) contributes to the attraction at short distances and slight Yukawa-like repulsion at long distances as elaborated in  \cite{Collins:2024sfi}
\begin{align}
V_\pi^{\rm cent}(&\vec q)=\left(\frac{g_c}{2f_\pi}\right)^2
\frac{\vec q^2}{q^2-m_\pi^2}
=\frac{g_c^2}{4f_\pi^2}\left(-1+\frac{\mu_\pi^2}{\vec q^2+\mu_\pi^2}\right),~\nonumber\\ \nonumber \\
 &V_\pi^{\rm cent}(r)=\frac{g_c^2}{4f_\pi^2}\left(-\delta^{(3)}(\vec r)+\frac{\mu_\pi^2}{4\pi r}e^{-\mu_\pi r}\right),
\label{Vcent}
\end{align}
which will be important for the interpretation of our results. 

The contact potential $V_{CT}$ near the threshold is parametrized via a low-energy expansion with the two lowest terms for $l=0$, and one term for $l=1$ as in \cite{Meng:2023bmz}. The employed potential for $DD^*$ system in Figure \ref{fig:effpot}b with CMF momenta $\vec p$ and $\vec p^{\,\prime}$ is then  
\begin{align} 
\label{V}
    &V \left( \vec{p}, \vec{\epsilon}; \vec{p}^{\,\prime} , \vec{\epsilon}^{~\prime}  \right) = \biggl[\left( 2 c_0^s + 2 c_2^s (\vec{p}^2 + \vec{p}^{~\prime~2} ) \right) \left( \vec{\epsilon} \cdot \vec{\epsilon}^{~\prime *} \right) \\
  &  +  2 c_2^p  \left( \vec{p}\cdot \vec{\epsilon} \right) \left( \vec{p}^{~\prime} \cdot \vec{\epsilon}^{~\prime *} \right)  
    +3 \bigg( \frac{g}{2 f_{\pi}} \bigg)^{2} \frac{\left( \vec{\epsilon} \cdot \vec{q} \right) \left( \vec{\epsilon}^{~\prime *} \cdot \vec{q} \right)}{q^2 - m_{\pi}^2} \biggr]\cdot f_{reg},\nonumber
\end{align}
which is illustrated in Figure~\ref{fig:effpot}b.  The last term accounts for the left-hand cut and incorporates it in our search for the pole of the scattering amplitude.  Additionally, three low-energy constants (LECs) are introduced: $c_{0,2}^s$ for s-wave and $c_2^p$ for p-wave. 

The function 
\begin{equation}
\label{freg}
f_{reg}(|\vec p|,|\vec p^{~\prime}|)=\exp \bigg(-\frac{|\vec p|^n+|\vec p^{~\prime }|^n}{\Lambda^n} \bigg)
\end{equation}
regularizes the potential, and our main result is based on a rather sharp cut-off with $n=40$ and $\Lambda=0.65 ~$GeV set near the $D^*D^*$ threshold, slightly above the energy of $D(1)D^*(-1)$ on our smaller volume.  Our main conclusions remain robust with the choices $n=10-40$ and $\Lambda=0.5-0.65~$GeV considered in the Appendix, while increasing $\Lambda$ further  is not appropriate due to the omission
of the $D^*D^*$ channel in the scattering analysis. 

\begin{figure*}[t!]
    \centering
    %\hspace{-1.4cm}
    \includegraphics[width=0.49\textwidth]{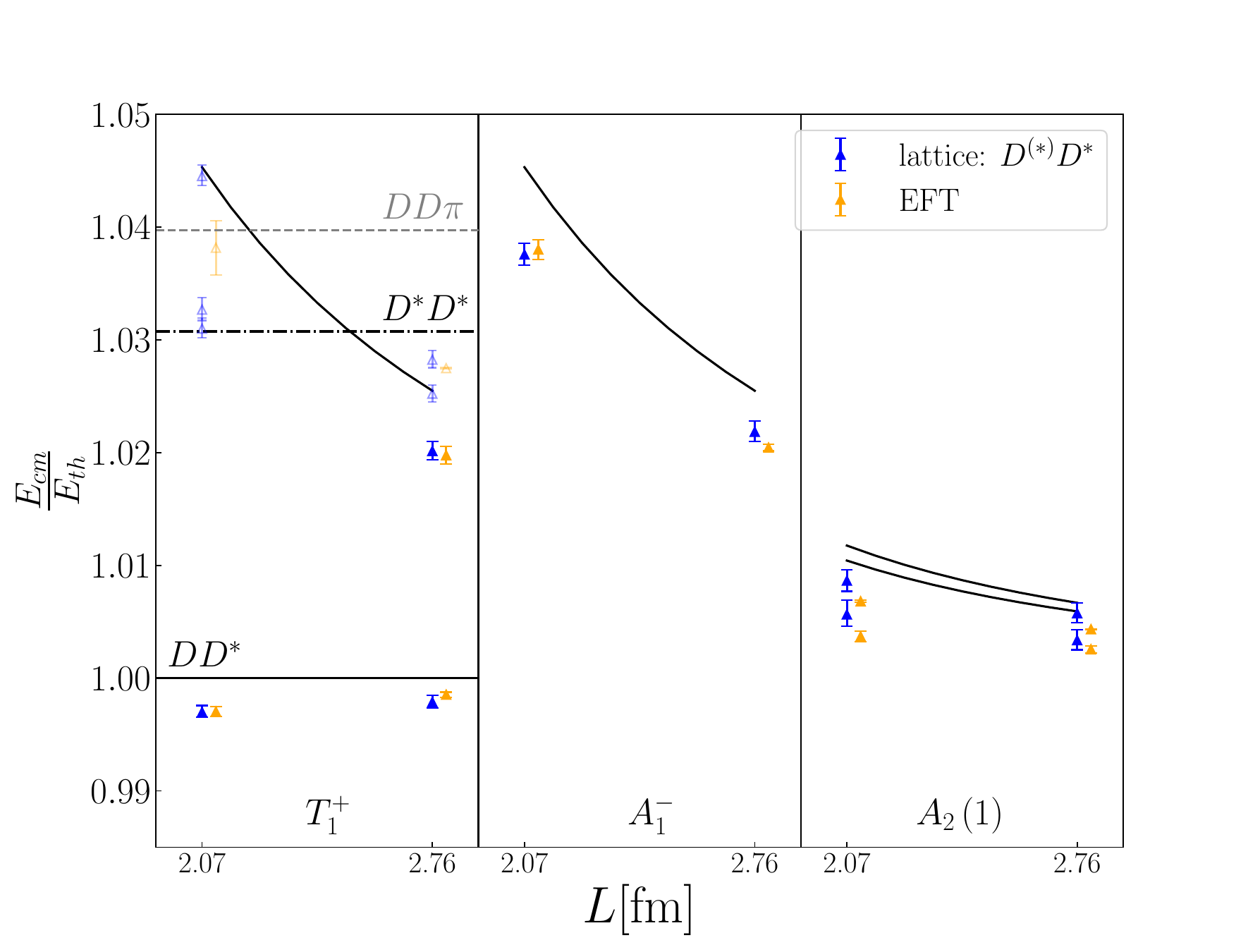}
    \includegraphics[width=0.49\textwidth]{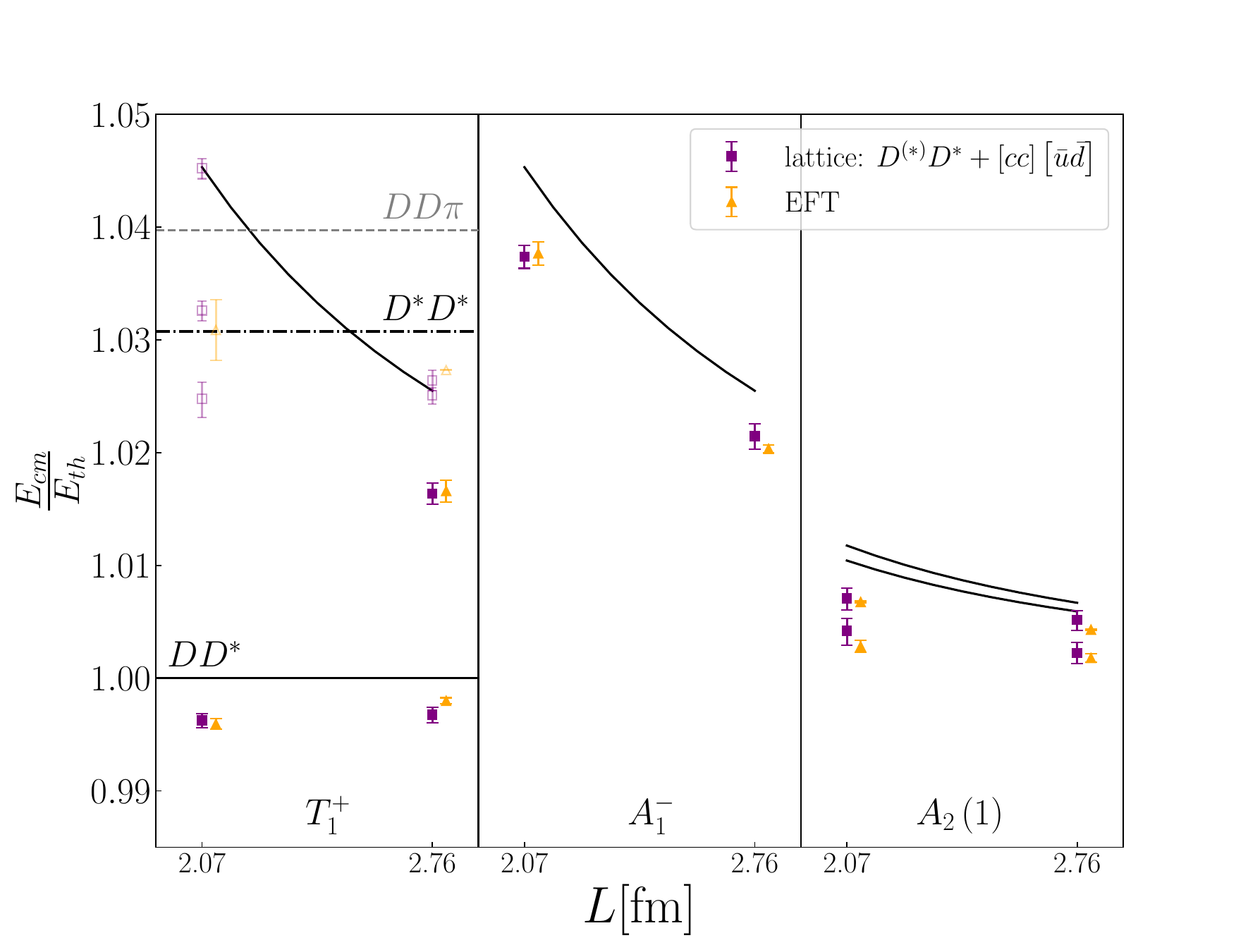}
    \caption{Fit of the low-energy constants parameterizing the potential: blue and violet points represent energies obtained from the lattice, while orange points are reconstructed from the fitted effective potential \eqref{eq:hameqirr}. The lattice data on the right incorporates the diquark-antidiquark interpolator, while the lattice data on the left omits it. The fit is based on the cut-off parameters   $\Lambda=0.65~\mathrm{GeV}$ and $n=40$ (\ref{freg}).}
   \label{fig:lecfit}
\end{figure*}

\subsection{Hamiltonian in the plane wave basis}
The natural choice of basis  for the matrices  in the Lippmann-Schwinger equation is composed of plane waves, which in the laboratory  and center-of-momentum frames take on the following forms, respectively:
\begin{align}
\label{eq:labpw}
    &| D(\vec{p}_D); ~ D^*(\vec p_{D^*}  ,\vec \epsilon^{~r}) \rangle_{lat},\ \ \vec P=\vec p_{D}+\vec p_{D^*}, \nonumber \\ \nonumber \\
    \qquad &\vec{p}_{D^{(*)}}\!\! \ = \ \!\! \tfrac{2\pi}{L}\vec{n}_{D^{(*)}}, \ \vec{n}_{D^{(*)}} \in Z^3; \ r = x, y, z, 
\end{align} 
and
\begin{align}
\label{eq:cmfpw}
    \hspace{0.62cm}| D (\vec{k} ); ~ D^*(-\vec k ,  \vec \epsilon^{~r} )\rangle_{cm} .
\end{align}
Our aim now is to evaluate the Hamiltonian $H$  in the plane wave basis and transform it to irreducible representation $ H^\Gamma$, where an example of this is explicitly shown in Appendix \ref{app:pw}. The plane-wave basis is finite due to the cut-off of the effective field theory, which is implemented in the potential via the regularization function $f_{reg}$.   
The expression (\ref{V}) for the potential $V$ applies in the CMF, therefore one forms the basis \eqref{eq:cmfpw}, which is obtained by a Lorentz transformation from the lattice frame to CMF\footnote{This Lorentz transformation does not modify polarization $\epsilon^r$  in the non-relativistic limit.}. The Hamiltonian matrix in plane wave basis is composed of matrix elements 
%\begin{align*}
%H_{\vec kr ,\vec k'r'} = _{cm}\!\langle D(\vec{k});  D^*(-\vec %k,  \vec \epsilon^{~r})|  H  | D(\vec{k}^{\prime});  D^*(-%\vec{k}^{\prime} ,  \vec \epsilon^{~r\prime} )\rangle_{cm} ,
%\end{align*}
\begin{align}
H_{\vec kr ,\vec k'r'} = \!\langle D(\vec{k});  D^*(-\vec k,  \vec \epsilon^{~r})|  H  | D(\vec{k}^{\prime});  D^*(-\vec{k}^{\prime} ,  \vec \epsilon^{~r\prime} )\rangle ,
\end{align}
with $H$ and $V$ defined in \eqref{eq:hameq} and (\ref{V}), respectively. The final step is applying a unitary transformation $U^\Gamma$, which converts the total plane-wave basis \eqref{eq:labpw} into a basis that transforms irreducibly with respect to the lattice symmetry group, where the irreducible representations are set to $\Gamma=T_1^+,~A_1^-,~A_2$. The resulting basis resembles the $DD^*$ operators in \eqref{ops}, and the projection technique to get this basis is well-established and explained in e.g. \cite{Meng:2021uhz, Prelovsek:2016iyo}. The Hamiltonian matrix $H^\Gamma$ in the irreducible basis $\Gamma$ then equals $ H^\Gamma=U^\Gamma HU^{\Gamma \dagger}$. Its energy spectrum (i.e. the eigenvalues of ${H}^\Gamma$) is afterward fitted to the observed lattice eigen-energies $E^{cm}$ obtained from the principal correlators $\lambda_n(t)$ shown in \eqref{gevp}.  

\section{Results on $\mathbf{DD^*} $ scattering}

This section provides results for the $DD^*$ potential, on-shell scattering amplitude, and the location of the $T_{cc}$ pole obtained from the lattice energies, following the formalism described in the previous section.

\subsection{Potential and   its low-energy constants  }
\label{subsec:potlec}
 The potential $V$ (\ref{V}) incorporates $s$-wave and $p$-wave interactions, and depends on three unknown low-energy constants $c_0^s, ~c_2^s$ and $c_2^p$. These are determined from the fit to the lattice eigen-energies. In particular, the eigen-energies of the effective Hamiltonian $H$ \eqref{eq:hameq} based on this potential and plane-wave basis are fitted to the lattice energies $E_{cm}$ indicated by filled symbols in Figure  \ref{fig:en_Z_c}. The resulting fits are shown in Figure \ref{fig:lecfit} and Table \ref{tab:lecfit} for the simulations including and excluding diquark-antidiquark operators. The reproduction of the lattice energies is particularly favorable for the data incorporating $[cc][\bar u\bar d]$ interpolators  ($\chi^2 / n_{dof}=1.4$), while also the fit omitting these operators is acceptable ($\chi^2 / n_{dof}=2.4$).    The inclusion of diquark-antidiquark operators mainly decreases the slope coefficient $c_2^s$, while the other two parameters remain unmodified.  
 These results are based on the preferred choice of a rather sharp cut-off $\Lambda\simeq 0.65~$GeV set near the $D^*D^*$ threshold with  $n=40$ in (\ref{freg}). Choices $\Lambda=0.5 \ \mathrm{GeV},~0.65~$GeV and $n=10-40$ lead to somewhat less favorable fits, while the main conclusions remain, as detailed in Appendix \ref{app:cutoff}.

\begin{figure*}[t!]
    %\hspace{-1.4cm}
    \centering
    \includegraphics[width=0.49\textwidth]{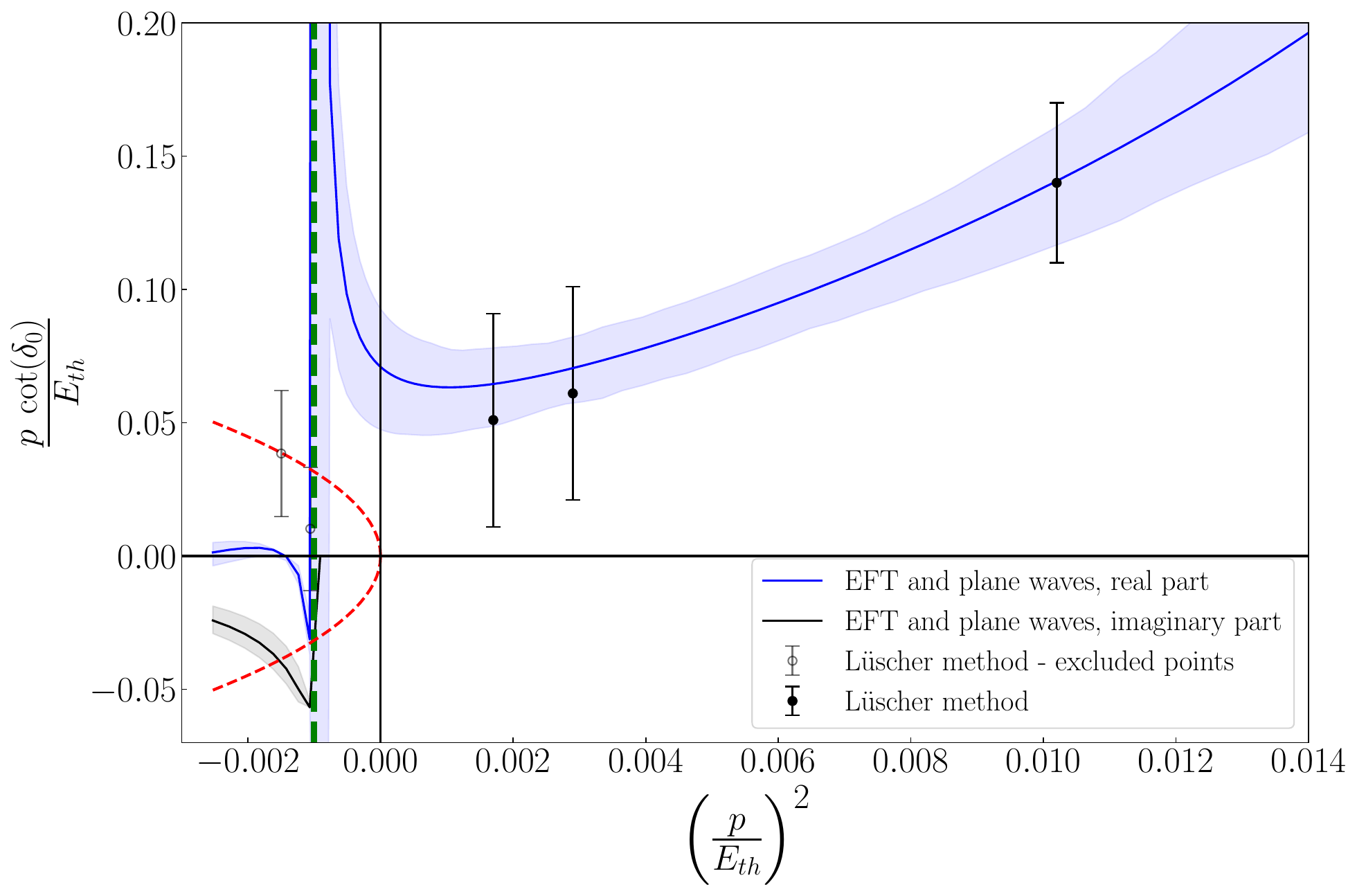}
    \includegraphics[width=0.49\textwidth]{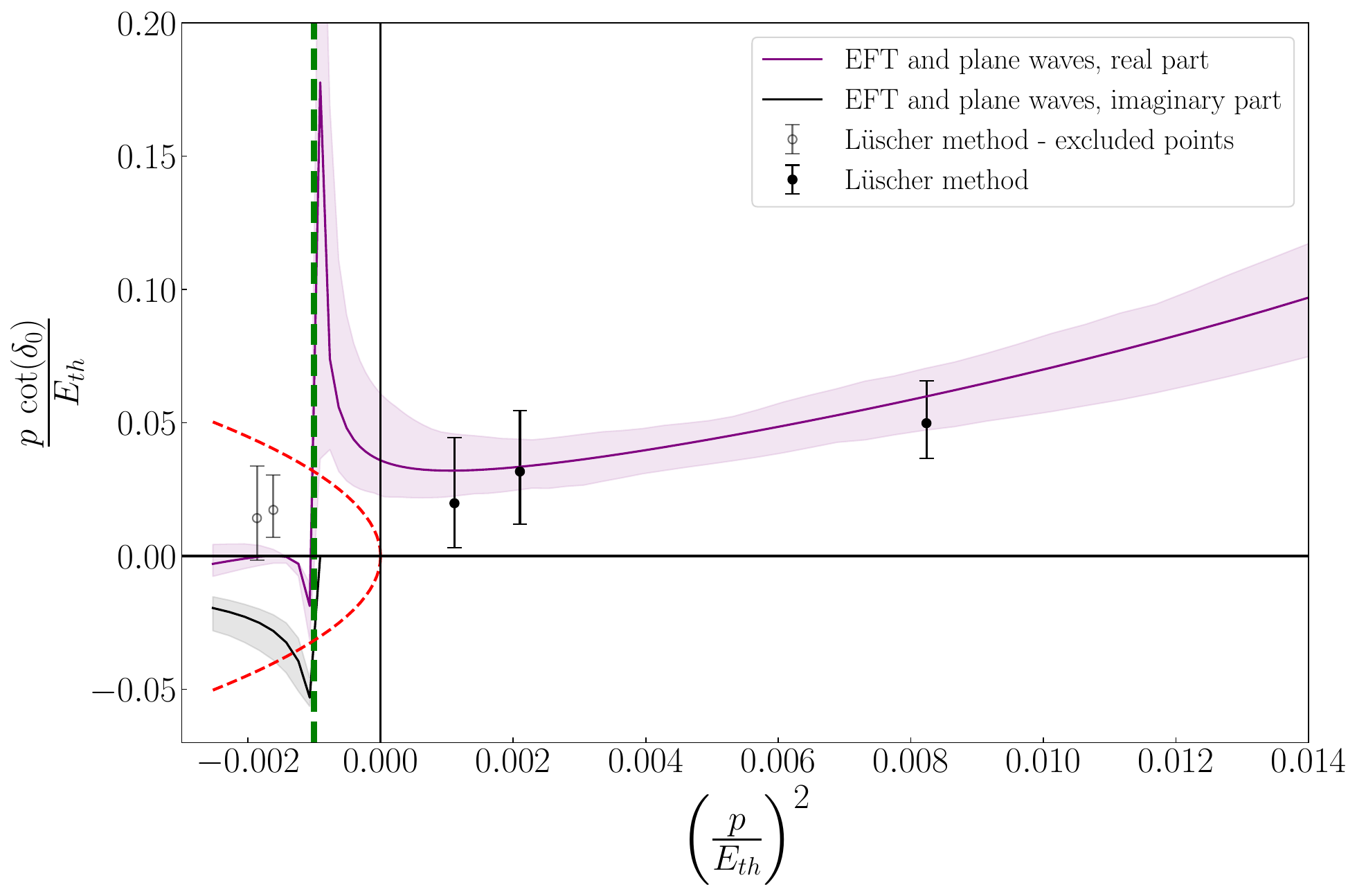}
    \caption{Comparison of $s$-wave $DD^*$ scattering phase shifts shown in terms of $p\cot \delta_0$ normalized to $E_{th}$: the left plot features only meson-meson data, while the analysis displayed in the right plot also incorporates a diquark-antidiquark interpolator.  The blue and violet bands represent $\mathrm{Re}(p\cot \delta_0)$  based on the EFT approach with  $s$ and $p$ waves, while the full and empty black circles are based on L\" uscher's method assuming negligible $p-$wave.  
    The vertical green line marks the beginning of the left-hand cut, below which the  EFT also renders an imaginary part of $p\cot \delta_0$  shown in gray. The red line represents $i p =\pm |p|$.}
    \label{fig:eftluscher}
\end{figure*}

\begin{table} 
\begin{center}
\begin{tabular}{c|c|c}
\hline
\hline
 & operators & operators\\ 
& $D^{(*)}D^*$ & $D^{(*)}D^*+[cc][\bar u\bar d]$ \\
\hline
$c_0^s~$[GeV$^{-2}$] & $-3.5\pm 1.0$ &  $-3.6\pm 1.0$   \\
$c_2^s~$[GeV$^{-4}$]& $6.6\pm 2.9$ & $1.3 \pm 2.9$ \\
$c_2^p~$ [GeV$^{-4}$]& $10.1\pm 1.0$  &  $9.8\pm 1.1$\\
$\chi^2/{\mathrm{dof}}$  & 2.4   & 1.4 \\
\hline
 & & \\
$E^p-E_{th}~$[MeV] & $-8.5^{+1.8}_{-2.4} \pm i \cdot 10.3^{+3.2}_{-4.1}$ &$-5.2^{+0.7}_{-0.8} \pm i \cdot 6.3^{+2.4}_{-4.8}$  \\
 & & \\
$E_{lhc}-E_{th}~$[MeV] & -7.98(5)& -7.98(5)\\ 
$(p_{lhc}/E_{th})^2\cdot 10^4$ & -10.03(8) & -10.03(8)\\
 \end{tabular}
\end{center}
\caption{The low-energy constants appearing in the $DD^*$ potential (\ref{V})  and $\chi^2$   from the fit of lattice eigen-energies based on EFT with cut-off $\Lambda=0.65~$GeV and $n=40$ (\ref{freg}). The $E^p$ represents the location of the pole of the resulting $DD^*$ scattering amplitude on the second Riemann sheet.   }\label{tab:lecfit} 
\end{table}

\subsection{Scattering amplitudes in  EFT plane-wave approach and Lüscher's approach}

 Once the low-energy constants and, therefore, the potential is fixed, the same potential is used in the infinite-volume Lippmann-Schwinger equation \eqref{eq:LSE} to determine the scattering amplitude. The scattering amplitude is related to the scattering phase shift  via \eqref{amplT} and the resulting 
 $p\cot  \delta_0 $ for s-wave is shown in Figure \ref{fig:eftluscher}. The values of $p\cot \delta_0$ near the threshold obtained with and without diquark-antidiquark operators are roughly similar: the inclusion of the diquark-antidiquark operators 
 shifts the value down by about 1-1.5 $\sigma$, which brings it closer to the crossing with the $ip=\pm |p|$ curve shown in red.  The inclusion of these operators also decreases the slope of  $p\cot \delta_0$ above the threshold, leading to smaller values of  $p\cot \delta_0$ and, therefore, a more attractive interaction.     Note that this   $s-$wave scattering result is obtained from lattice energies using a fit that incorporates $s-$ as well as  $p-$wave interactions between $D$ and $D^*$. 

  The same  Figure \ref{fig:eftluscher} also shows the values of $p\cot  \delta_0 $ based on L\"uscher's method, where the black circles are obtained from individual energy levels assuming a negligible $p-$wave interaction.  In line with expectations, both methods are in good agreement at energies above the left-hand cut, marked with a vertical green line. 
  The two approaches are not expected to agree in the vicinity of or below the left-hand cut since the EFT incorporates this cut while L\"uscher's approach omits it.  

\begin{figure}[t!]
    \centering
    \includegraphics[width=0.5\textwidth]{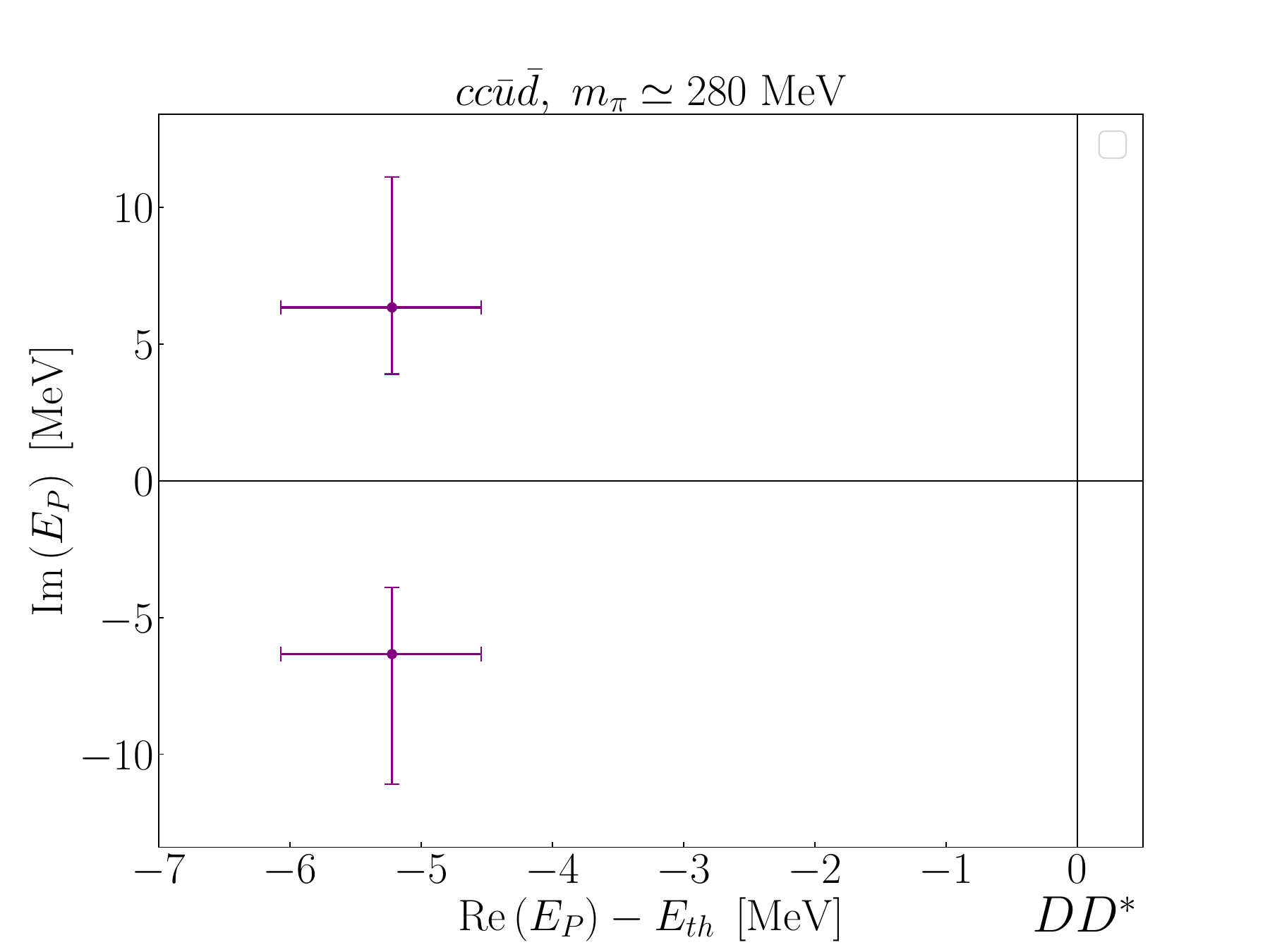}
    \caption{The resulting location of the $T_{cc}^{+}$ pole in $DD^*$ scattering amplitude at the employed $m_\pi\simeq 280~$MeV.  The pole appears on the second Riemann sheet, and the origin represents the $DD^*$ threshold. This result is based on the full operator basis. }
    \label{fig:tccpole-final}
\end{figure}

\begin{figure}[t!]
    \centering
    \includegraphics[width=0.5\textwidth]{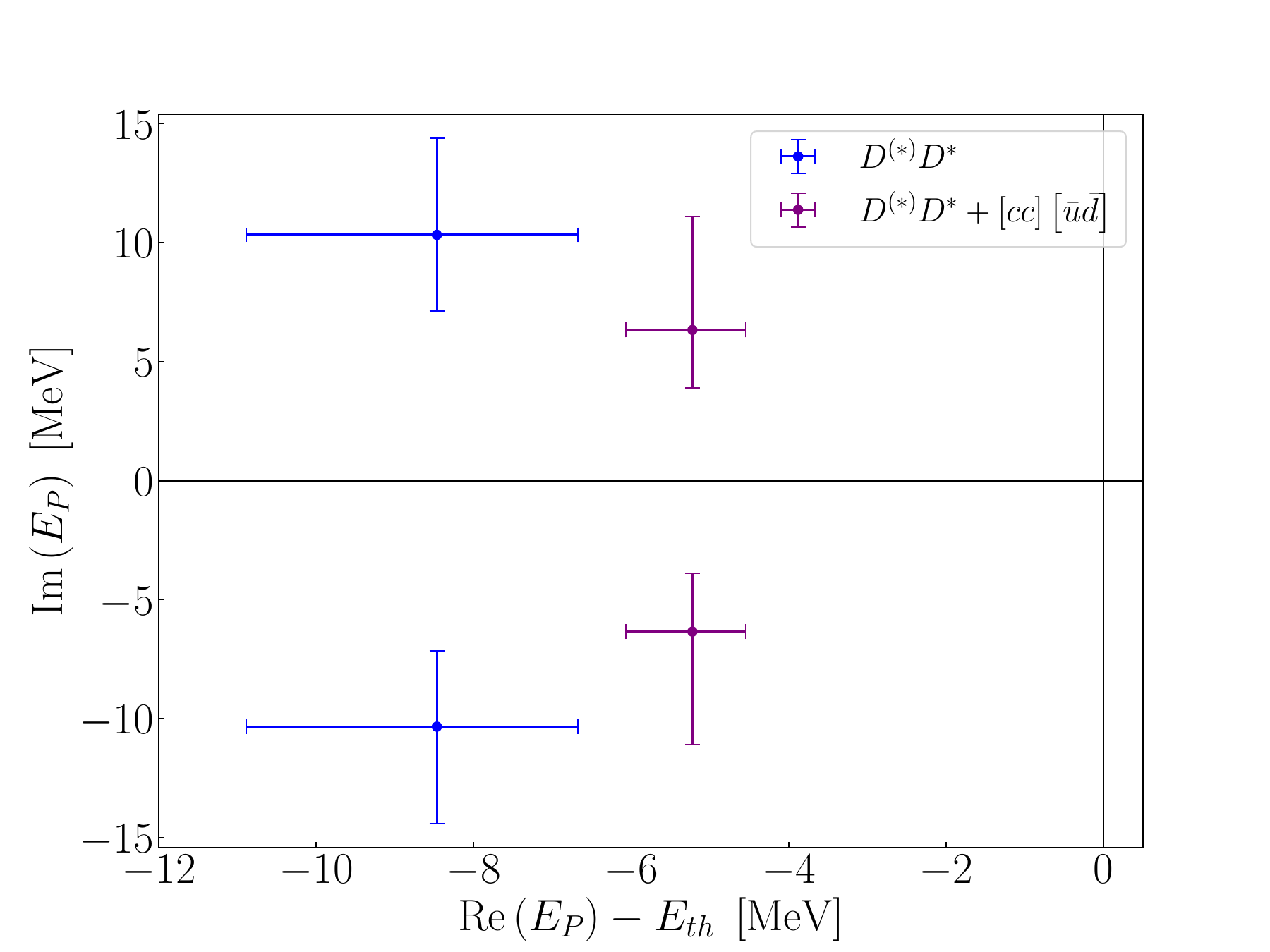}
    \caption{The locations of the $T_{cc}^{+}$ poles in $DD^*$ scattering amplitude, shown also in Table \ref{tab:lecfit}. The violet pair of points encompass all operators listed in Section \ref{operator_basis} and are also shown in the previous Fig. \ref{fig:tccpole-final}. The blue points result from the simulation that excludes the diquark-antidiquark operator.}
    \label{fig:tccpole}
\end{figure}

\subsection{\texorpdfstring{$T_{cc}^{+}$}{Tccplus} pole}

In order to search for poles, the scattering amplitude is continued to the complex energy plane and  Riemann sheets I and II are explored\footnote{Riemann sheet $I$ corresponds to $\mathrm{Im}(p)>0$, while $II$ corresponds to $\mathrm{Im}(p)<0$. }.  The corresponding on-shell scattering amplitude $T_I$ for complex $p$ and $E=E_{th}+p^2/2m_r$  is a solution of Lippmann-Schwinger equation \eqref{eq:LSE}, where the integral runs over the real momenta  $d\vec k$. The amplitude on the second sheet is obtained via $T_{II}^{-1}=T_I^{-1}-2i\frac{m_r}{2\pi}\sqrt{2m_r E}$ in order to satisfy $S_{II}(E)=S_{I}^{-1}(E)$\footnote{Here $S_{I\atop II}=1-2i\frac{m_r}{2\pi}p_{I\atop II}T_{I\atop II}$ and the square-root function has a cut on the positive real axes with $\mathrm{Im(E)}\geq 0$. }.  

The pole location of  $T_{cc}^{+}$  is presented in Figure \ref{fig:tccpole-final} and Table \ref{tab:lecfit}. This result is based on our full operator basis and the pole appears as a subthreshold resonance on the second Riemman sheet.    

The comparison of pole $T_{cc}^{+}$ locations for the operator basis, including and excluding diquark-antidiquark operators, is shown in Figure \ref{fig:tccpole}  and Table \ref{tab:lecfit}.  In both cases, the tetraquark appears as a subthreshold resonance, which corresponds to a pole on the second Riemann sheet that lies less than 10 MeV below the $DD^*$ threshold.  The  
data that incorporates the diquark-antidiquark interpolator shows somewhat greater attraction in the system. The inclusion of this operator shifts the pole slightly closer to the threshold and the physical scattering axis where the pole would turn to a virtual state pole. This is consistent with the $p\cot\delta_0$ curve approaching $ip=|p|$ in Figure \ref{fig:eftluscher}; the crossing of $p\cot\delta_0$ and $ip=-|p|$ would imply a virtual state pole at real energy below the threshold. 

\subsection{ Discussion}
Our lattice simulation considers $DD^*$ scattering in a kinematical situation where $D^*$ is stable since $m_\pi\simeq 280~$MeV$>m_{D^*}-m_D$.  The attraction in the $DD^*$ system and thereby the presence of the pole follows from the attractive $DD^*$ potential at short distance, represented by the negative contact term $c_0$ (\ref{V}) and negative term in the one-pion exchange (\ref{Vcent}). Focusing on $s-$wave interaction, the fully attractive potential would have rendered a virtual or a bound state pole on the real energy axes. 
However, the $DD^*$ scattering at lattice  kinematics with $m_\pi >m_{D^*}-m_D$ receives a contribution from one-pion 
exchange   (\ref{Vcent}) that renders also a slightly repulsive Yukawa-like part at longer distances. This slightly repulsive part is responsible for $T_{cc}$ featuring as a sub-threshold resonance at physical charm quark mass and our $m_\pi $. 

One anticipates that with decreasing $m_\pi$ and/or increasing heavy quark mass, the $T_{cc}$ resonance pole will transition to the virtual-state pole and then to the bound state pole, as elaborated in \cite{Collins:2024sfi,Gil-Dominguez:2024zmr,Abolnikov:2024key} based on the existing lattice simulations. 
 
\subsection{Outlook}

\indent This paper explores doubly heavy tetraquarks by analyzing their finite-volume energy spectrum using diquark-antidiquark operators at two different heavy quark masses. It addresses the left-hand cut in the $DD^*$ scattering amplitude with the use of an effective potential evaluated in the plane-wave basis. However, this approach represents only one out of several that have been developed recently to deal with these issues.  \\
\indent One could employ the relativistic version of combined EFT and plane-wave approach  \cite{Meng:2021uhz, Meng:2023bmz}, besides the non-relativistic version that was used here due to the proximity of considered lattice energies to the $DD^*$ threshold.\\
\indent Various alternative ways of dealing with the left-hand cuts in scattering amplitudes have been proposed. For example, in Ref. \cite{Raposo:2023oru} the two-particle L\"uscher's formalism is generalized to explicitly account for the left-hand cut by relaxing some requirements of the original quantization condition, while more recently this has been extended to coupled channels and arbitrary spin in \cite{Raposo:2025dkb}. Ref. \cite{Bubna:2024izx} proposed a modified quantization condition in the presence of long-range forces that usually arise due to exchanges of light particles, such as one-pion exchange explicitly treated in this work, while Ref. \cite{Dawid:2024oey} provides another strategy to deal with these challenges. 
 In addition to these, Refs. \cite{Hansen:2014eka, Dawid:2024dgy} develop and apply to the $T_{cc}$, respectively, the 3-body quantization condition that effectively extends its range of validity all the way up to the first 4-particle states that can go on shell in a given channel. In this formalism $D^*$ features as $D\pi$ bound state when $m_\pi>m_{D^*}-m_D$. \\
\indent Another interesting possibility for a future study would be applying approach used here to investigate the pole trajectory of doubly-charmed tetraquark as a function of light and heavy quark masses. 
This was touched upon in this paper and examined in our previous paper  \cite{Collins:2024sfi}, albeit without the diquark-antidiquark operator in the basis. Dependence of the $T_{cc}$ pole on the masses of the constituent quarks was analyzed also in \cite{Gil-Dominguez:2024zmr, Abolnikov:2024key}. \\
\indent The present one-channel study could also benefit from the extension to the coupled-channel $DD^* - D^*D^*$. This has been recently done in \cite{Whyte:2024ihh} at $m_\pi\simeq 390~$MeV with an expanded operator basis. Therein the authors apply L\"uscher's quantization condition to lattice energies that are above the left-hand cut located below $DD^*$ threshold. 

In \cite{Stump:2024lqx} a novel method of implementing diquark-antidiquark operators is proposed that is based on position-space sampling, thereby circumventing computational costs that arise naturally within distillation, as explained in more detail in section \ref{sec:dist}, already at a relatively modest number of Laplacian eigenvectors. This method could be used in future simulations involving the diquark-antidiquark operator.

\section{Conclusions  } 
This work presented lattice QCD results on doubly heavy tetraquarks $QQ\bar u\bar d$ with $J^P\!=\!1^+$ and $I\!=\!0$ for $m_\pi\!\simeq\! 280~$MeV and heavy quark $Q$ with charm or close to bottom quark mass.  

Building upon the already existing meson-meson bilocal interpolators, we implemented additional localized diquark-antidiquark interpolators with the distillation method and explored their effects. We find that for $Q=c$ the diquark-antidiquark operator has a somewhat small $1-2~\sigma$ effect on certain eigen-energies, the scattering amplitude, and the resulting pole location 
within our simulation framework. 
For $Q\simeq b$, the $BB^*$ scattering operators alone do not render a deeply bound $T_{bb}$, and the inclusion of diquark-antidiquark operators is required, which shifts the ground state energy significantly down while also influencing other energy levels. 

Our study presents the first extraction of $DD^*$ scattering amplitude based on meson-meson as well as diquark antidiquark operators. 
The  scattering amplitude was extracted from lattice eigen-energies in a framework combining an effective field theory and plane-wave approach. This framework is applicable also for energies on the left-hand cut, which is present in the lattice kinematics with stable $D^*$. Three low-energy constants of the EFT potential were fitted from nine lattice energies leading to a favorable reproduction of the lattice data.  The resulting scattering amplitude agrees with the one obtained with the L\" uscher's approach in the energy region above the left-hand cut where the latter is applicable. 

The $T_{cc}$ is found as a subthreshold resonance corresponding to a pole at   $m_{T_{cc}}-m_D-m_{D^*}=-5.2^{+0.7}_{-0.8} - i \cdot 6.3^{+2.4}_{-4.8}~$MeV at the employed $m_\pi\!\simeq\! 280~$MeV. The presence of the pole near threshold results from a significant attractive interaction between $D$ and $D^*$.  
A small shift of the pole away from the real axes can be traced back to a slightly repulsive part of the one-pion exchange interaction at larger distances in our kinematics where $m_\pi> m_{D^*}-m_D$. Note that in the LHCb experiment, the $T_{cc}$ pole is away from the real axes due to the three-body decay  $DD\pi$, which is kinematically closed in our simulation as well as for all other existing lattice simulations.

\section{Acknowledgments}
We would like to thank V. Baru, S. Dawid, E. Epelbaum, L. Meng, A. Nefediev, F. Romero-López, and S. Sharpe for illuminating discussions. We thank our colleagues in CLS for the joint effort in the generation of the gauge field ensembles which form a basis for the computation. We use the multigrid solver of Refs.~\cite{Heybrock:2014iga,Heybrock:2015kpy,Richtmann:2016kcq,Georg:2017diz} for the inversion of the Dirac operator. Our code implementing distillation is written within the framework of the Chroma software package \cite{Edwards:2004sx} and utilizes the Eigen library \cite{eigenweb}. The authors gratefully acknowledge the HPC RIVR consortium (\href{https://www.hpc-rivr.si}{www.hpc-rivr.si}) and EuroHPC JU (\href{https://eurohpc-ju.europa.eu/}{eurohpc-ju.europa.eu}) for funding this research by providing computing resources of the HPC system Vega at the Institute of Information Science (\href{www.izum.si}{www.izum.si}). M.P. acknowledges the use of computing clusters at IMSc Chennai. We thank the authors of Ref.~\cite{Morningstar:2017spu} for making the \textit{TwoHadronsInBox} package public. The work of I. V., S. P. and L. L. is supported by the Slovenian Research Agency (research core Funding No. P1-0035 and J1-3034 and N1-0360). M.P. gratefully acknowledges support from the Department of Science and Technology, India, SERB Start-up Research Grant No. SRG/2023/001235 and Department of Atomic Energy, India.

\appendix

\section{Variation of cut-off }\label{app:cutoff}

 The $DD^*$ scattering amplitude in the main text is based on the effective field theory with our preferred values of the cut-off $\Lambda=0.65~$GeV and rather a sharp fall-off in the regulator $f_{reg}$  for the potential (\ref{V},\ref{freg}) obtained using $n=40$. Such a cut-off is slightly above the non-interacting level $D(1)D^*(-1)$ on our smaller volume. 

\begin{figure*}[t!]
    \centering
    \begin{subfigure}{0.49\textwidth}
        \includegraphics[width=\textwidth]{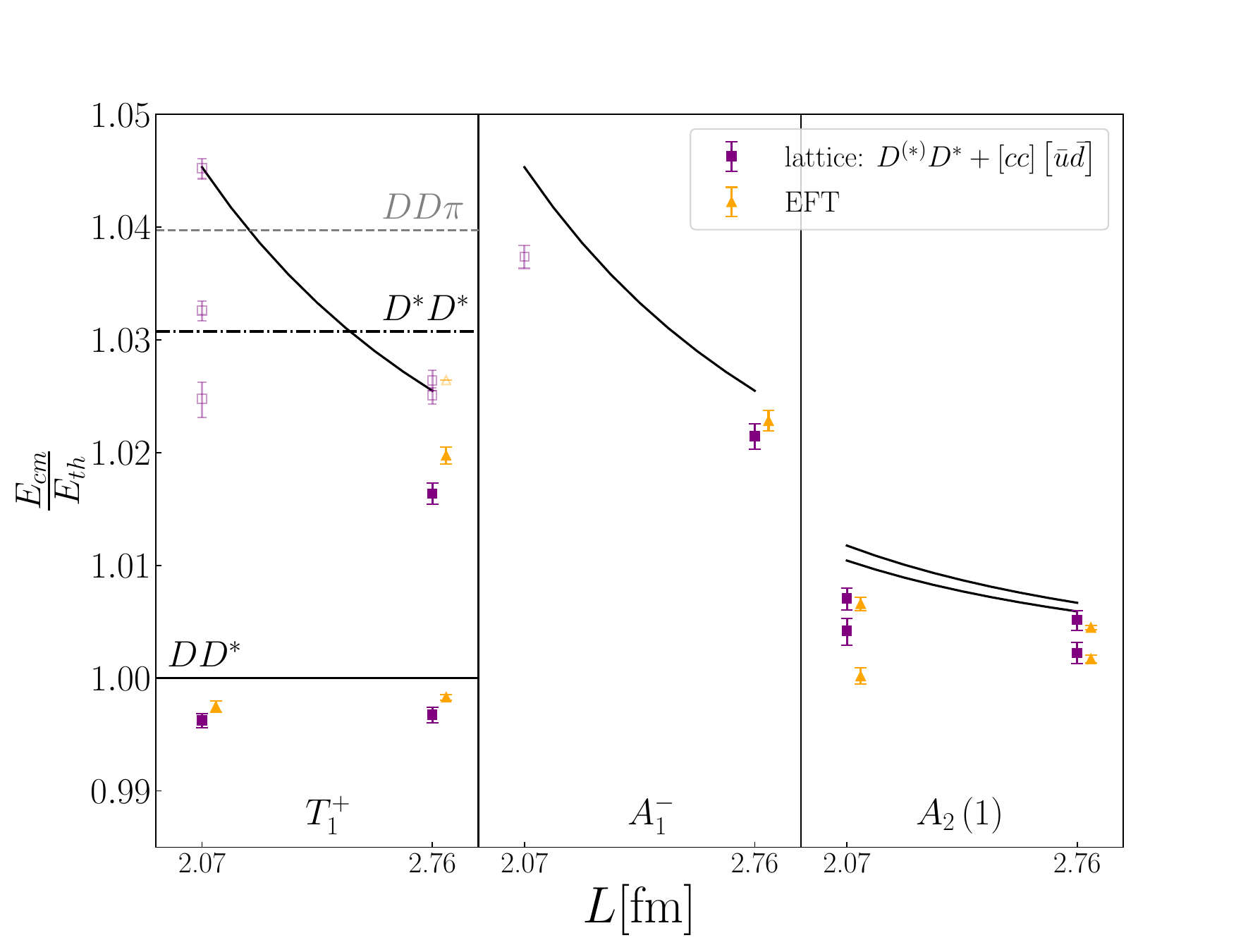}
        \caption{$\Lambda = 0.5~\mathrm{GeV},~n=10,~\frac{\chi^2}{n_{dof}} = 6.8$.}
    \end{subfigure}
    \begin{subfigure}{0.49\textwidth}
        \includegraphics[width=\textwidth]{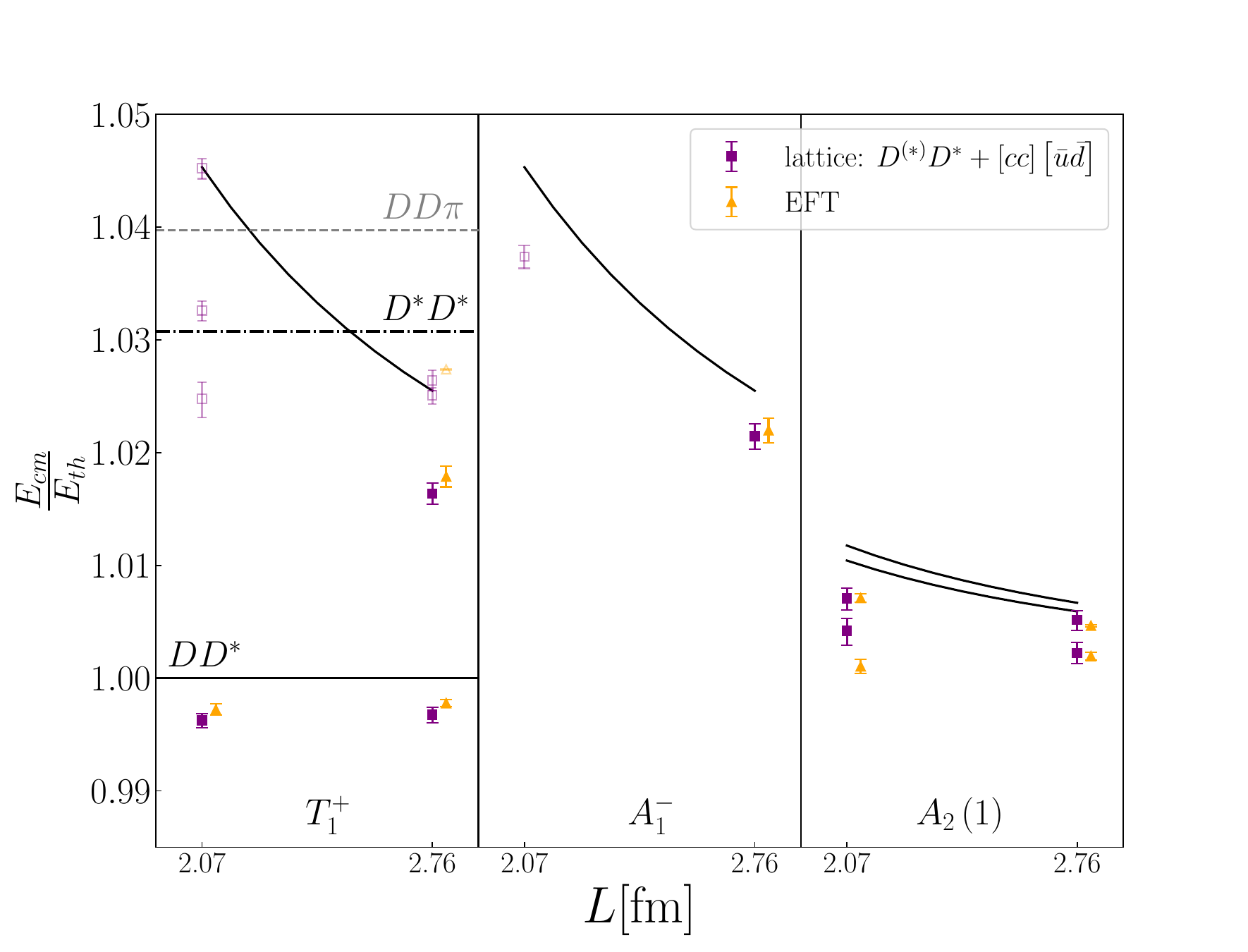}
        \caption{$\Lambda = 0.5~\mathrm{GeV},~n=40,~\frac{\chi^2}{n_{dof}} = 3.2$.}
    \end{subfigure}
    \vspace{0.5cm}
    \begin{subfigure}{0.49\textwidth}
        \includegraphics[width=\textwidth]{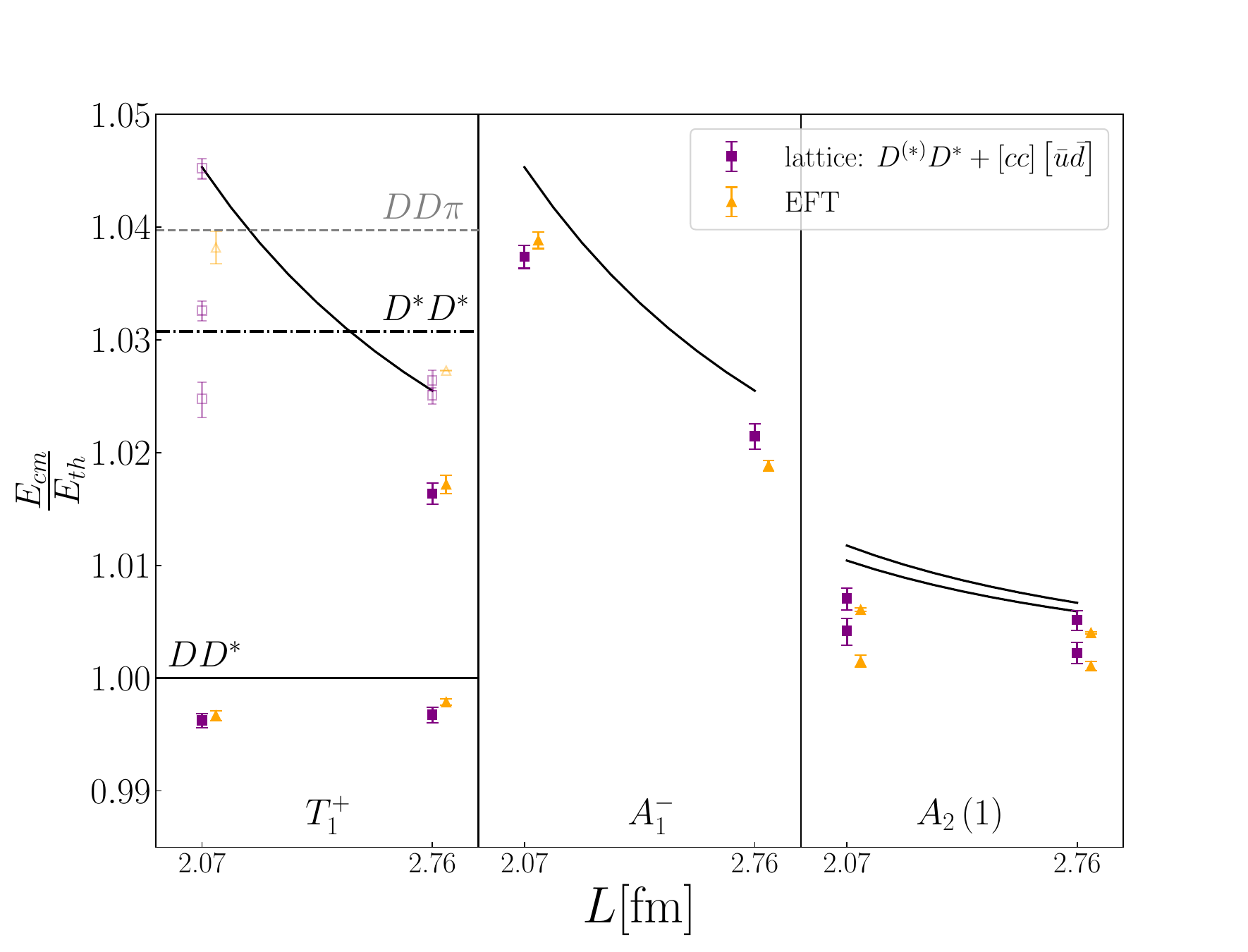}
        \caption{$\Lambda = 0.65~\mathrm{GeV},~n=10,~\frac{\chi^2}{n_{dof}} = 4.1$.}
    \end{subfigure}
    \begin{subfigure}{0.49\textwidth}
        \includegraphics[width=\textwidth]{MM4q_eft_lattice_0.65_40.pdf}
        \caption{$\Lambda = 0.65~\mathrm{GeV},~n=40,~\frac{\chi^2}{n_{dof}} = 1.3$.}
    \end{subfigure}
    \caption{Fit of the low-energy constants for various cut-offs $\Lambda$ and shapes $n$ in the regulator $f_{reg}$ (\ref{freg}) of the potential: violet points represent lattice energies obtained with all interpolators, while orange points are the energies reconstructed from the fitted effective potential.}
    \label{fig:lecfit-app}
\end{figure*}

\begin{figure*}[t!]
    \centering
    \begin{subfigure}{0.48\textwidth}
        \includegraphics[width=\textwidth]{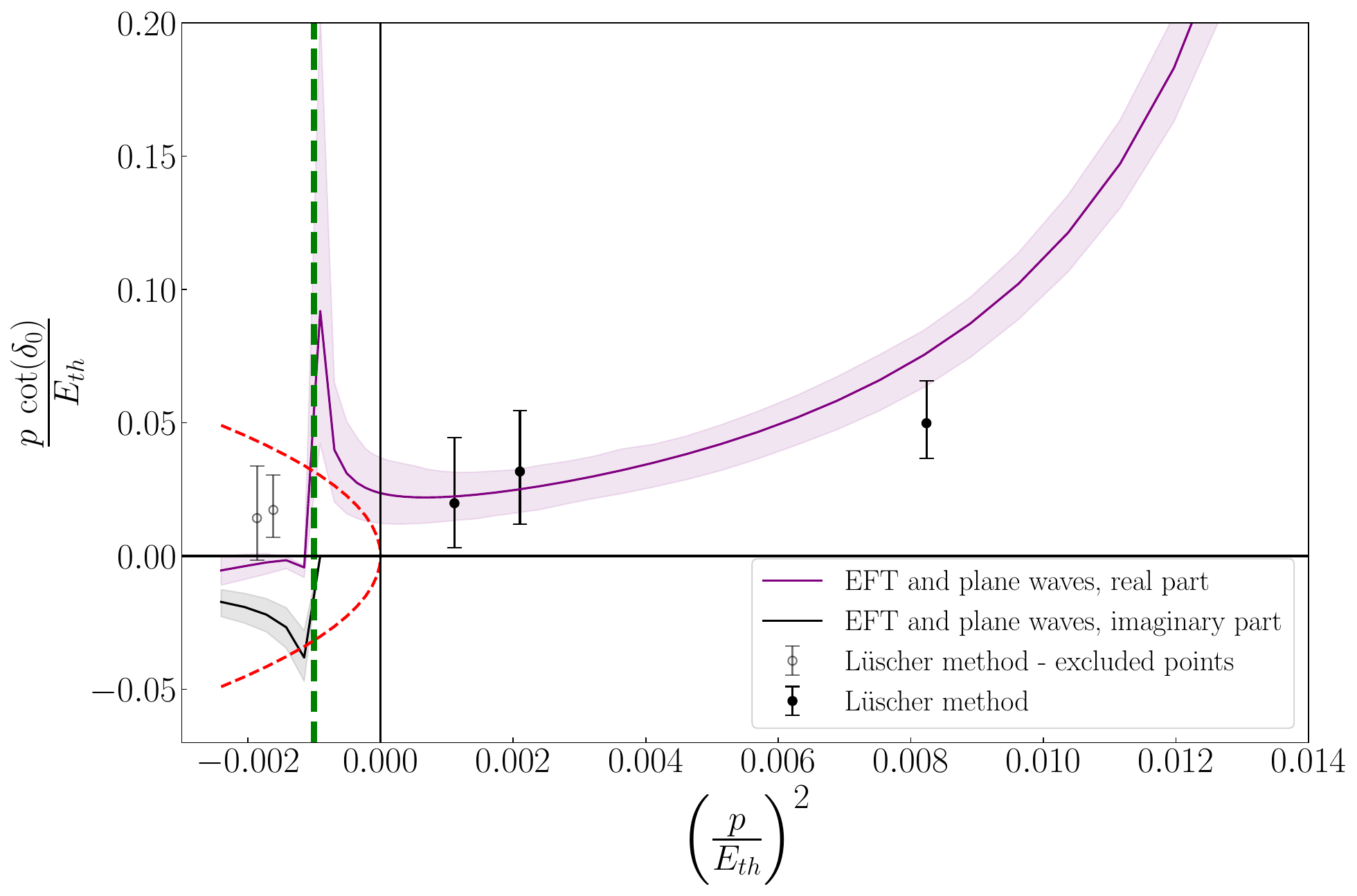}
        \caption{$\Lambda = 0.5~\mathrm{GeV},~n=10,~\frac{\chi^2}{n_{dof}} = 6.8$.}
    \end{subfigure}
    \begin{subfigure}{0.48\textwidth}
        \includegraphics[width=\textwidth]{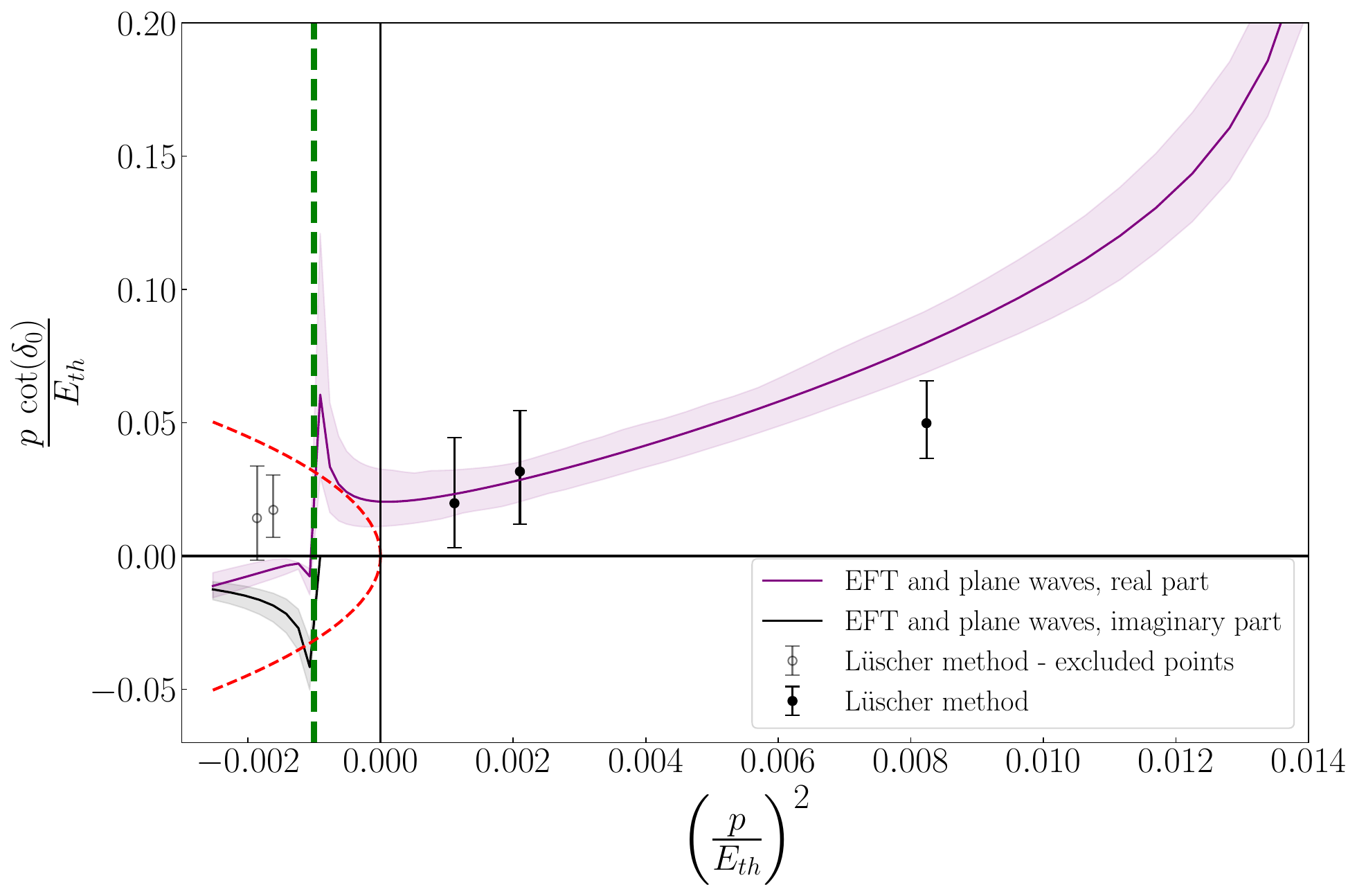}
        \caption{$\Lambda = 0.5~\mathrm{GeV},~n=40,~\frac{\chi^2}{n_{dof}} = 3.2$.}
    \end{subfigure}
    \vspace{0.5cm}
    \begin{subfigure}{0.48\textwidth}
        \includegraphics[width=\textwidth]{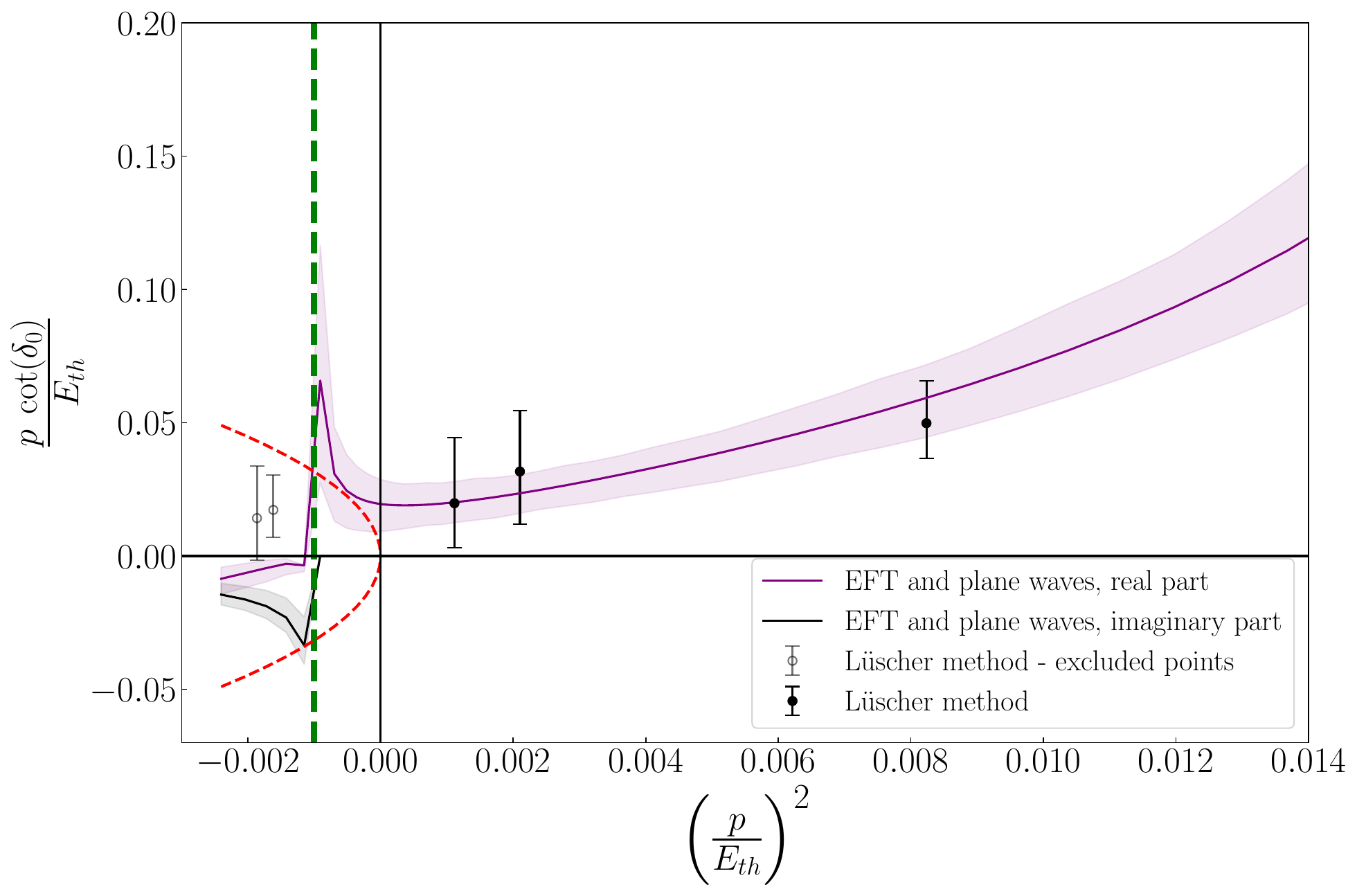}
        \caption{$\Lambda = 0.65~\mathrm{GeV},~n=10,~\frac{\chi^2}{n_{dof}} = 4.1$.}
    \end{subfigure}
    \begin{subfigure}{0.48\textwidth}
        \includegraphics[width=\textwidth]{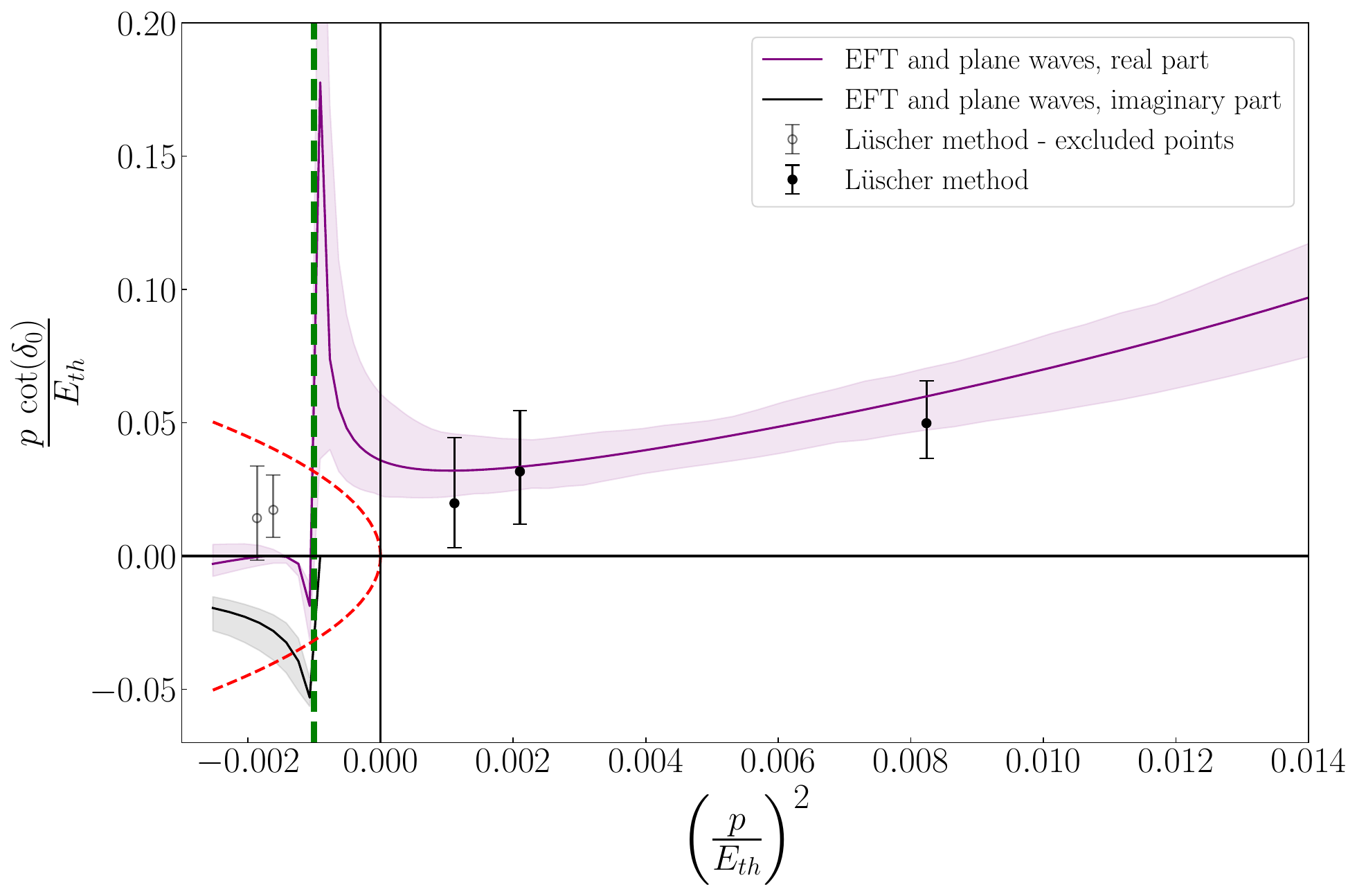}
        \caption{$\Lambda = 0.65~\mathrm{GeV},~n=40,~\frac{\chi^2}{n_{dof}} = 1.3$.}
    \end{subfigure}
    \caption{ $p\cot ( \delta_0 )$ for $DD^*$ scattering obtained using all interpolators. Violet curve:   The result is based on an EFT fit to lattice energies (with given $\chi^2$)  for various $\Lambda$ and $n$ in the potential regulator $f_{reg}$ (\ref{freg}). Black circles: results based on L\" uscher's approach which is applicable only above the left-hand cut shown by the green dashed line.   }
    \label{fig:eftluscher-MM4q-app}
\end{figure*}

\begin{figure*}[t!]
    \centering
    \begin{subfigure}{0.48\textwidth}
        \includegraphics[width=\textwidth]{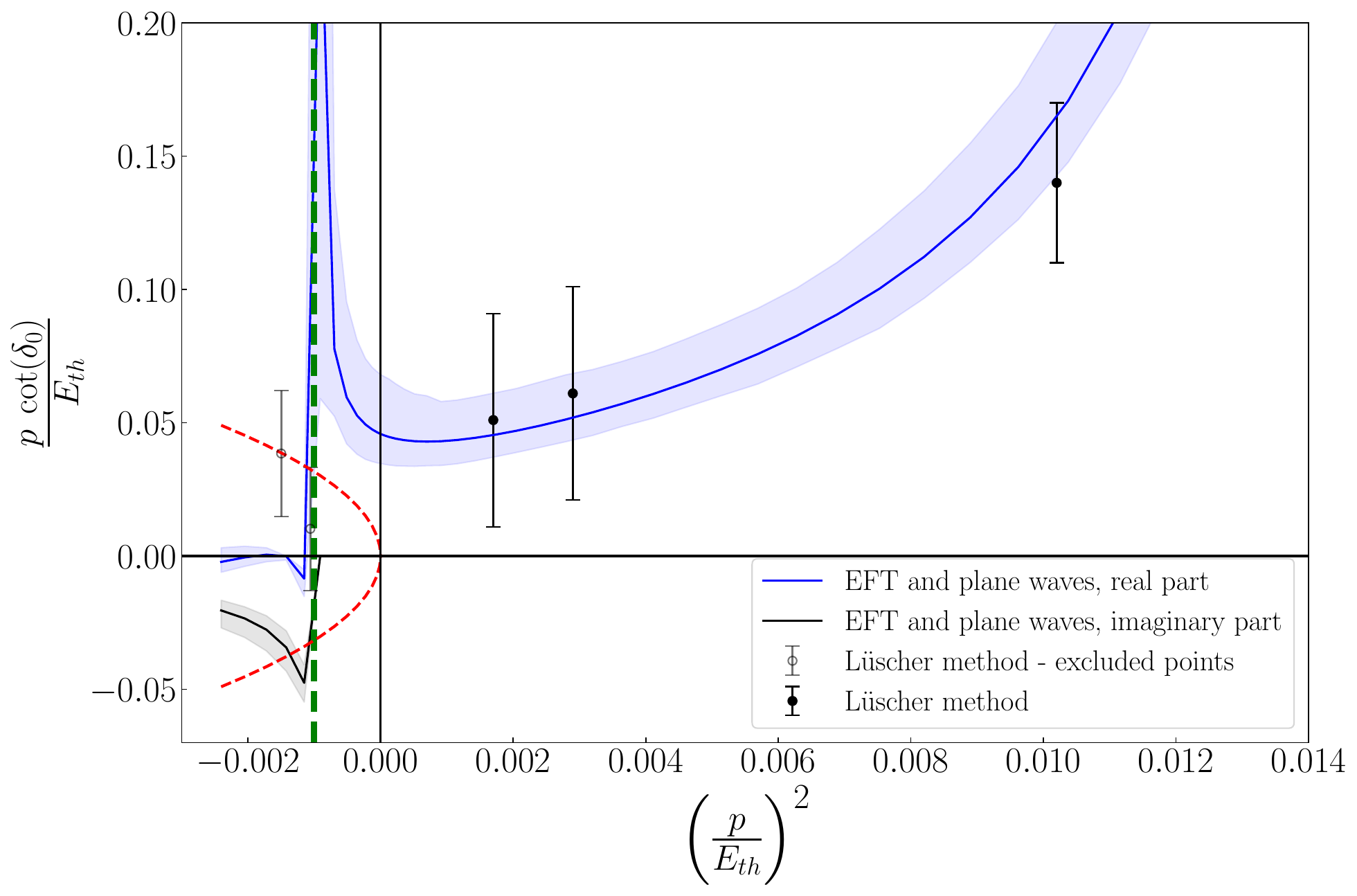}
        \caption{$\Lambda = 0.5~\mathrm{GeV},~n=10,~\frac{\chi^2}{n_{dof}} = 5.4$.}
    \end{subfigure}
    \begin{subfigure}{0.48\textwidth}
        \includegraphics[width=\textwidth]{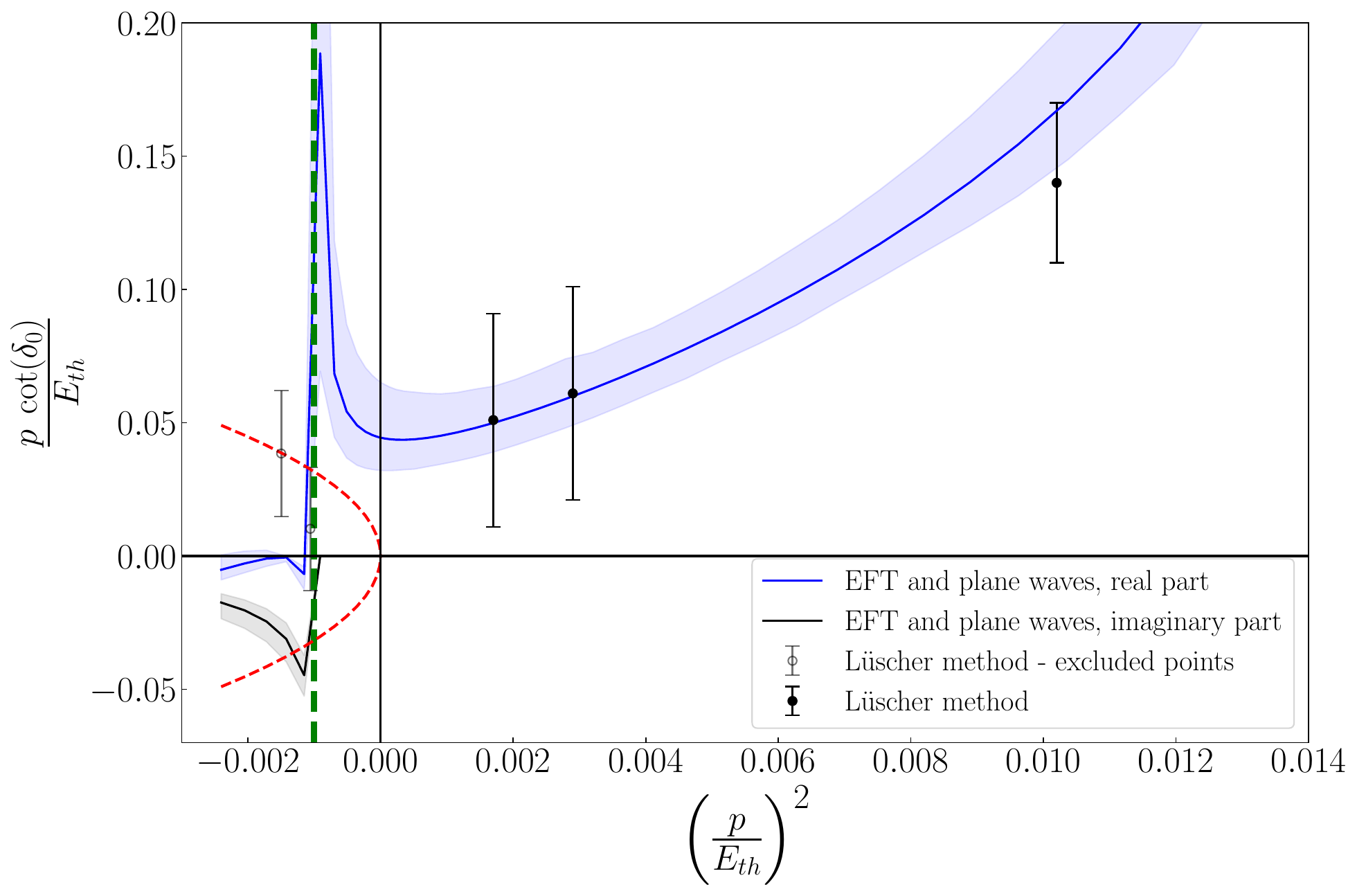}
        \caption{$\Lambda = 0.5~\mathrm{GeV},~n=40,~\frac{\chi^2}{n_{dof}} = 2.8$.}
    \end{subfigure}
    \vspace{0.5cm}
    \begin{subfigure}{0.48\textwidth}
        \includegraphics[width=\textwidth]{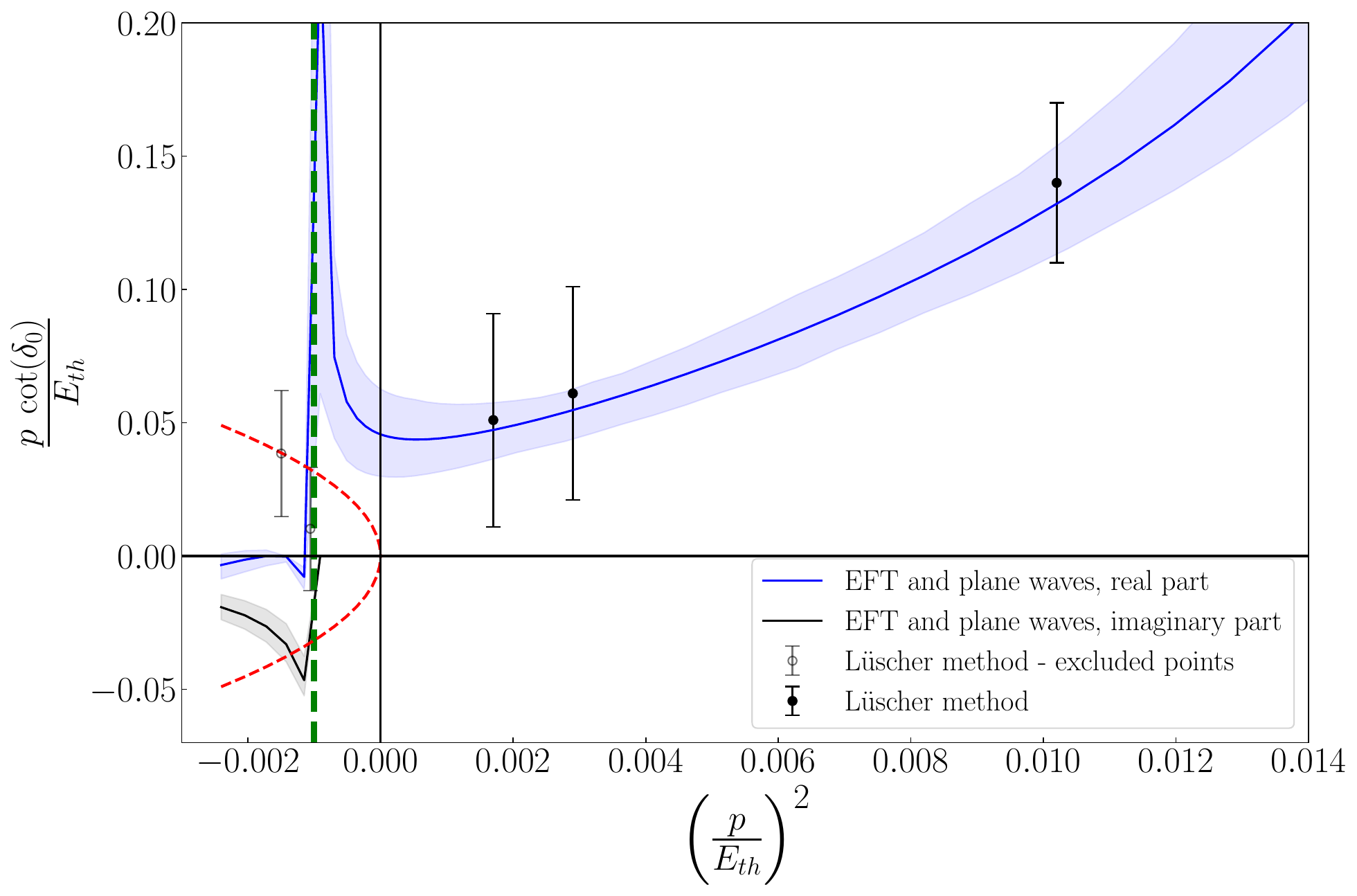}
        \caption{$\Lambda = 0.65~\mathrm{GeV},~n=10,~\frac{\chi^2}{n_{dof}} = 5.5$.}
    \end{subfigure}
    \begin{subfigure}{0.48\textwidth}
        \includegraphics[width=\textwidth]{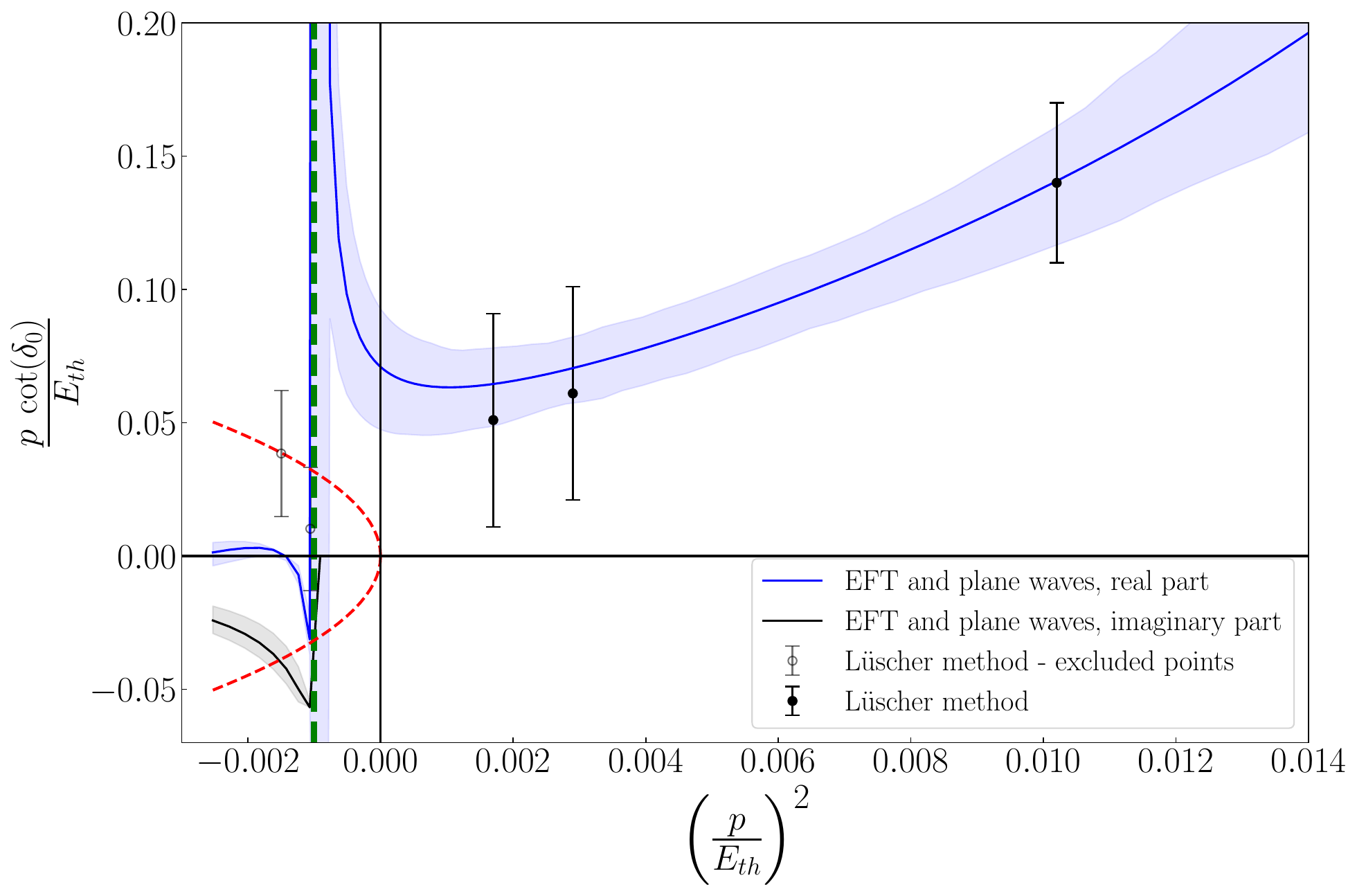}
        \caption{$\Lambda = 0.65~\mathrm{GeV},~n=40,~\frac{\chi^2}{n_{dof}} = 2.4$.}
    \end{subfigure}
    \caption{  Same as Figure \ref{fig:eftluscher-MM4q-app}, but for lattice data based solely on meson-meson operators. }
    \label{fig:eftluscher-MM-app}
\end{figure*}

\begin{figure*}[t!]
    \centering
    \begin{subfigure}{0.48\textwidth}
        \includegraphics[width=\textwidth]{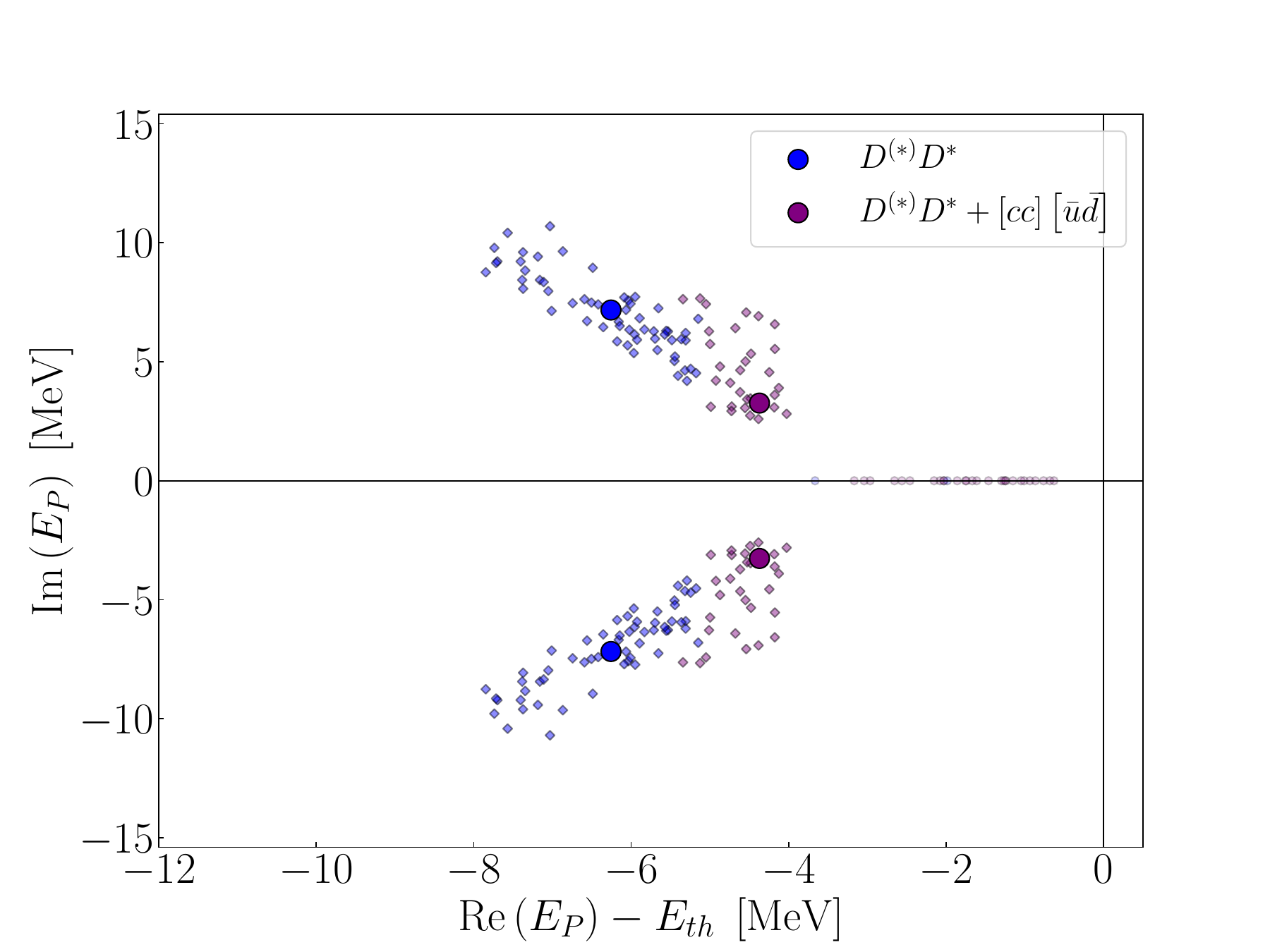}
        \caption{$\Lambda = 0.5~\mathrm{GeV},~n=10,~{\color{violet}\frac{\chi^2}{n_{dof}}=6.8},~{\color{blue}\frac{\chi^2}{n_{dof}}=5.4}$.}
    \end{subfigure}
    \begin{subfigure}{0.48\textwidth}
        \includegraphics[width=\textwidth]{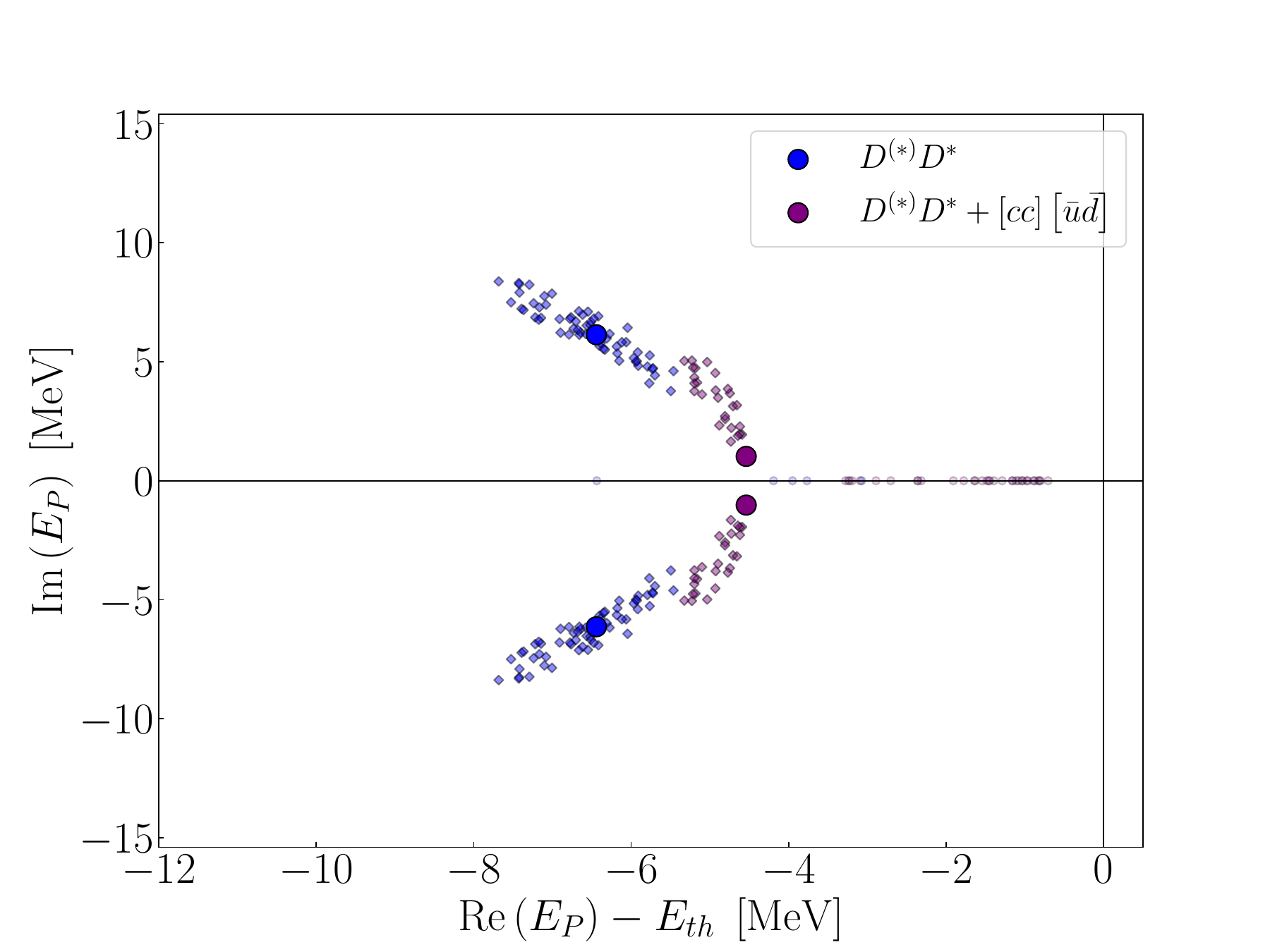}
        \caption{$\Lambda = 0.5~\mathrm{GeV},~n=40,~{\color{violet}\frac{\chi^2}{n_{dof}}=3.2},~{\color{blue}\frac{\chi^2}{n_{dof}}=2.8}$.}
    \end{subfigure}
    \vspace{0.5cm}
    \begin{subfigure}{0.48\textwidth}
        \includegraphics[width=\textwidth]{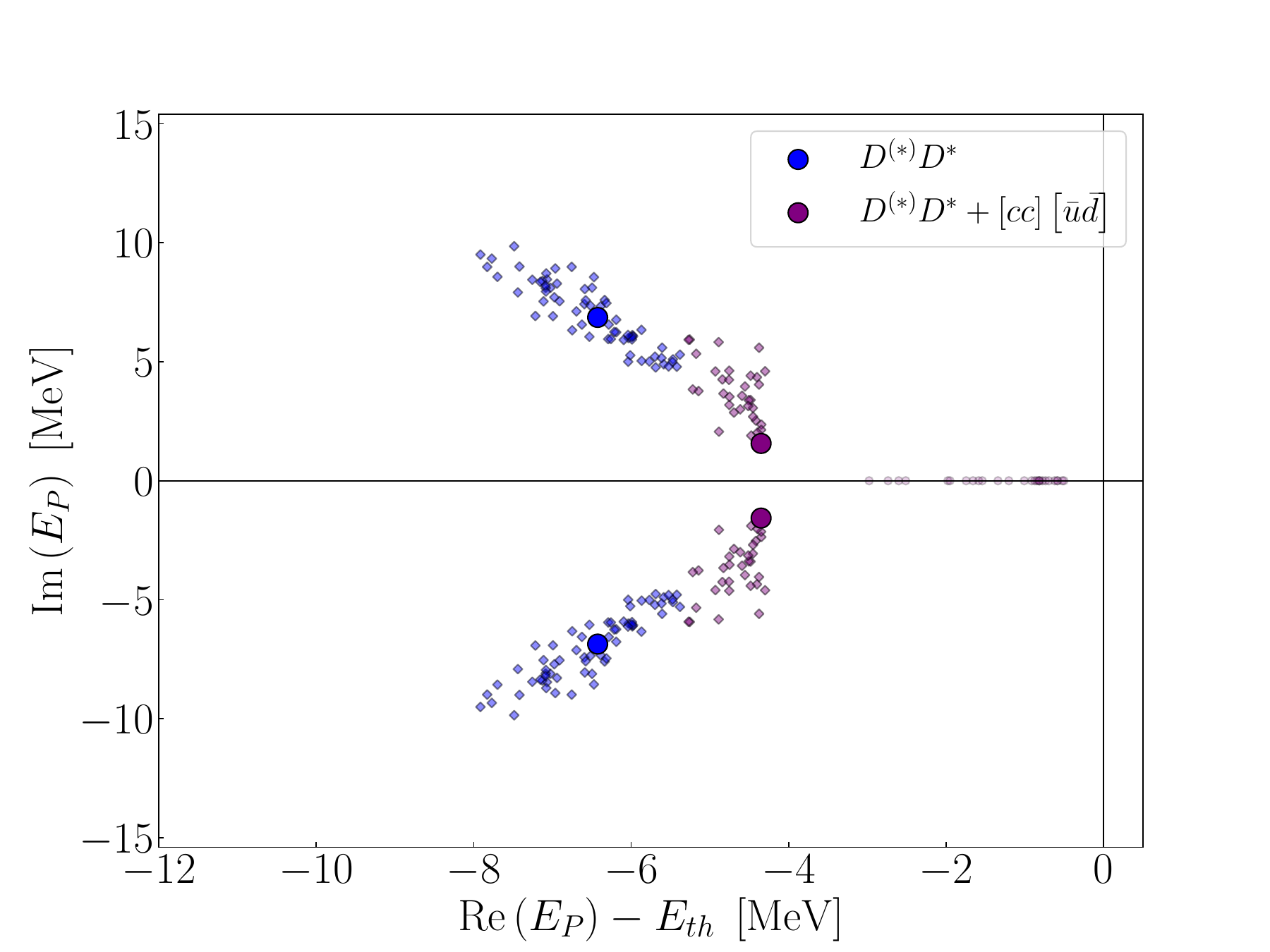}
        \caption{$\Lambda = 0.65~\mathrm{GeV},~n=10,~{\color{violet}\frac{\chi^2}{n_{dof}}=4.1},~{\color{blue}\frac{\chi^2}{n_{dof}}=5.5}$.}
    \end{subfigure}
    \begin{subfigure}{0.48\textwidth}
        \includegraphics[width=\textwidth]{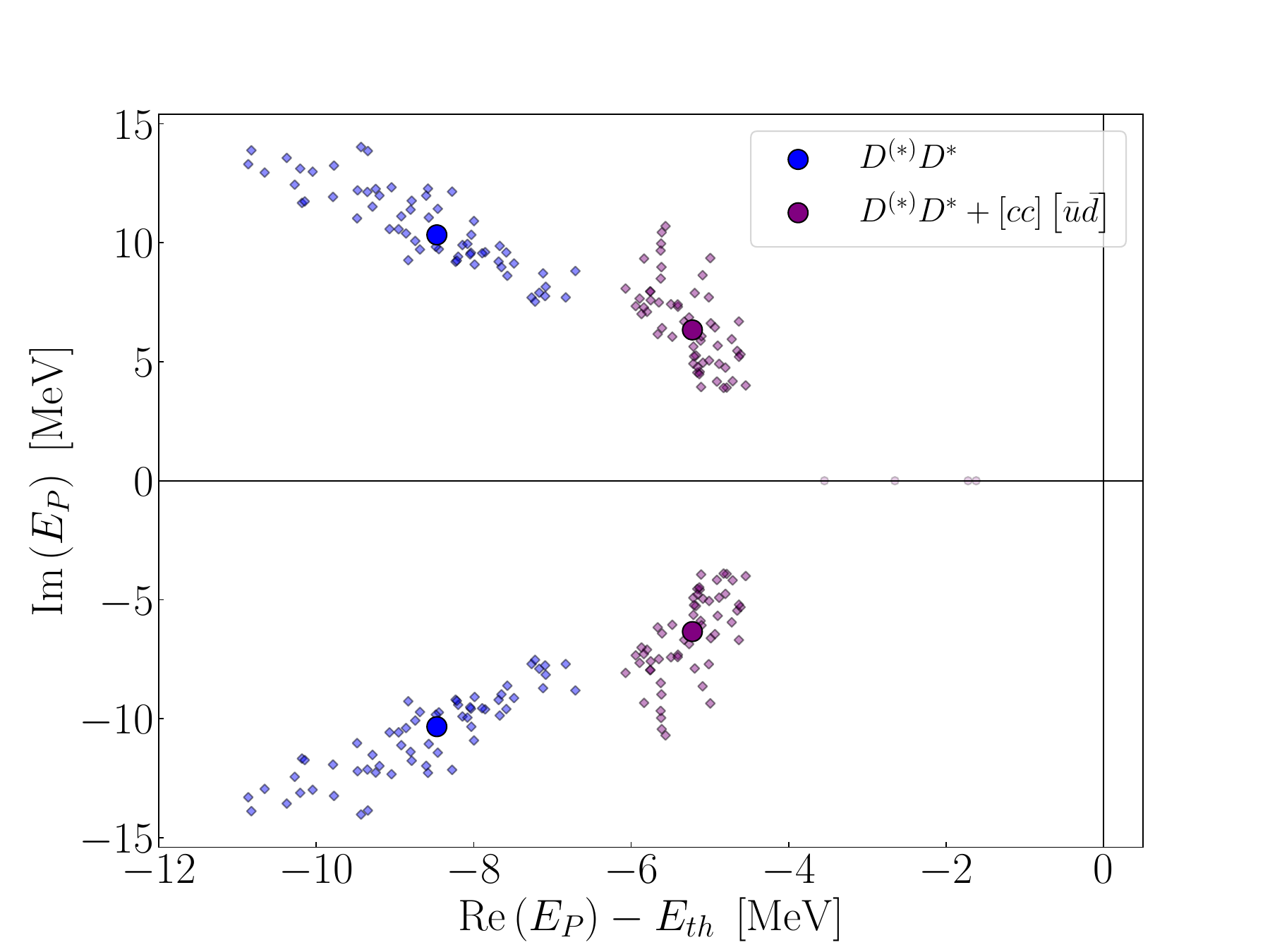}
        \caption{$\Lambda = 0.65~\mathrm{GeV},~n=40,~{\color{violet}\frac{\chi^2}{n_{dof}}=1.3},~{\color{blue}\frac{\chi^2}{n_{dof}}=2.4}$.}
    \end{subfigure}
    \caption{Location of the $T_{cc}^{+}$ pole obtained from our lattice results including (violet) or excluding (blue) diquark antidiquark operators. The results based on various cut-offs  $\Lambda$ and $n$ in the potential regulator $f_{reg}$ (\ref{freg}) are shown.  
    The large circle represents the central value, while the distribution of the small diamonds 
    represents the $1\sigma$ error band.   More precisely, the dispersed points shown are generated from a hundred pseudorandom samples of the low-energy constants that are normally distributed according to their central values and the covariance matrix obtained from the fit to the lattice spectrum, as explained in Subsection \ref{subsec:potlec}. Each of these points is calculated by taking one such sample and solving the LSE \eqref{eq:LSE}. The points outside of $1\sigma$ range of central values are not shown. The origin represents the $DD^{*}$ threshold.   
    }
    \label{fig:tccpole-app}
\end{figure*}

 This appendix presents the fits to lattice eigen-energies   (Figure \ref{fig:lecfit-app}), the corresponding $p\cot\delta_0$ (Figures \ref{fig:eftluscher-MM4q-app} and \ref{fig:eftluscher-MM-app}) and pole positions (Figure \ref{fig:tccpole-app}) also for a smoother regulator $n=10$ and lower cut-off $\Lambda=0.5~$GeV. The reproduction of certain lattice energies and resulting $\chi^2$ is less favorable, however, the main conclusions based on all these choices remain robust: (i) the inclusion of diquark-antidiquark operators somewhat decreases $p\cot\delta_0$ and moves the $T_{cc}$  pole slightly closer to the $DD^*$ threshold, (ii) the resulting $p\cot\delta_0$ based on EFT and L\" uscher's approach show reasonable agreement in the region above the left-hand cut, and (iii) the $T_{cc}$ is a sub-threshold resonance with a pole within $10~$MeV  of the threshold for all cases considered. 

\begin{figure*}[t!]
    \centering
    \begin{subfigure}{0.48\textwidth}
        \includegraphics[width=\textwidth]{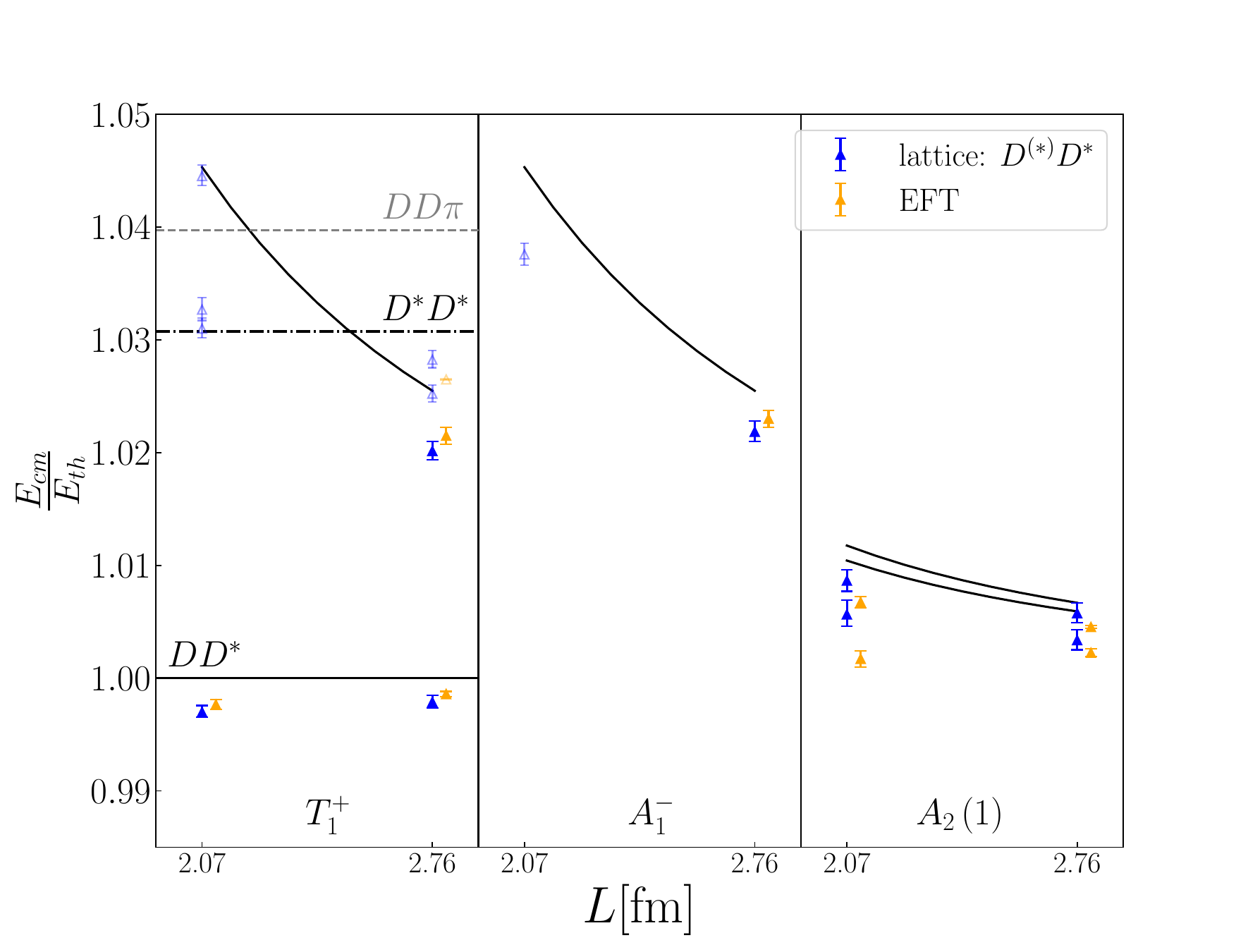}
        \caption{$\Lambda = 0.5~\mathrm{GeV},~n=10,~\frac{\chi^2}{n_{dof}} = 5.4$.}
    \end{subfigure}
    \begin{subfigure}{0.48\textwidth}
        \includegraphics[width=\textwidth]{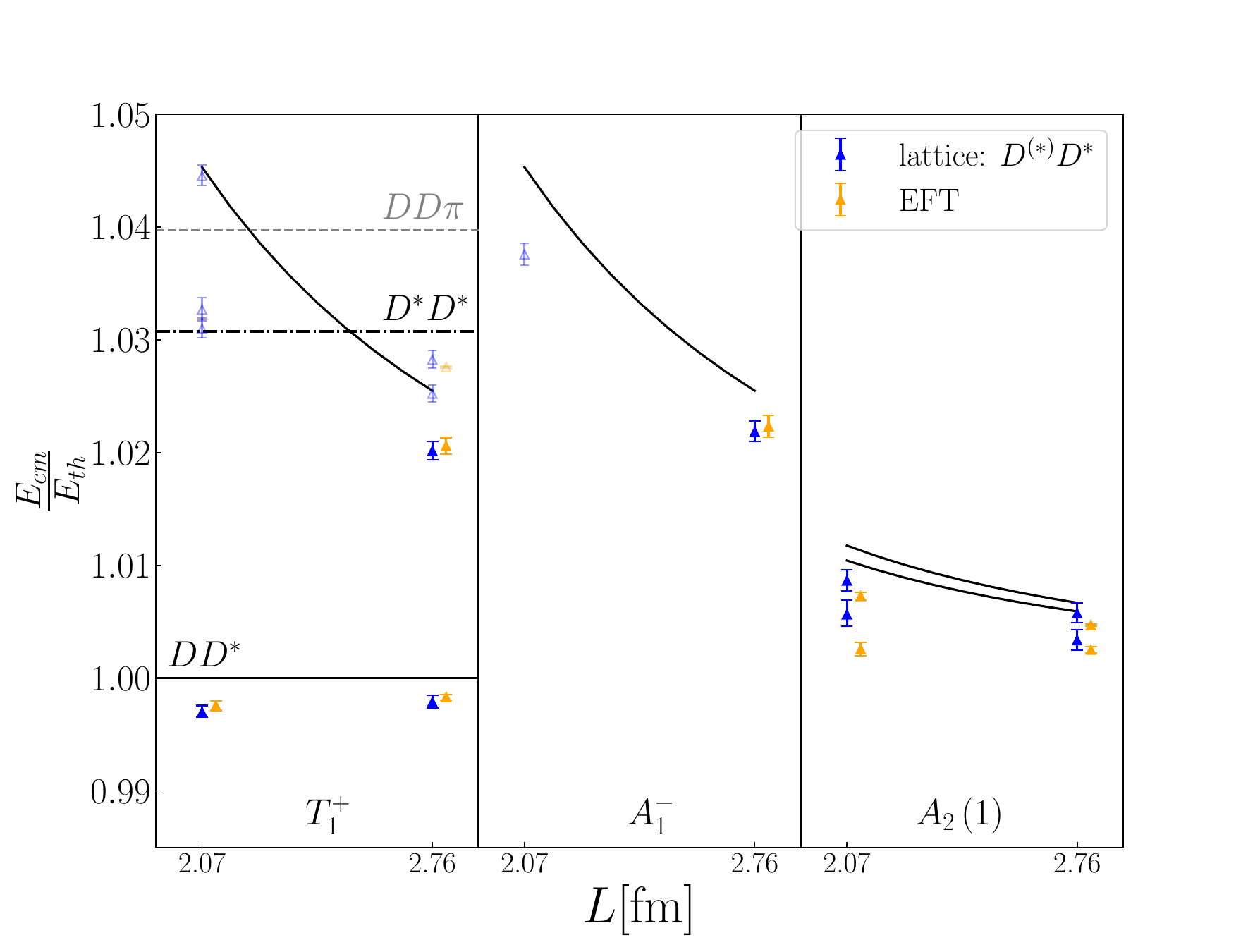}
        \caption{$\Lambda = 0.5~\mathrm{GeV},~n=40,~\frac{\chi^2}{n_{dof}} = 2.8$.}
    \end{subfigure}
    \vspace{0.5cm}
    \begin{subfigure}{0.48\textwidth}
        \includegraphics[width=\textwidth]{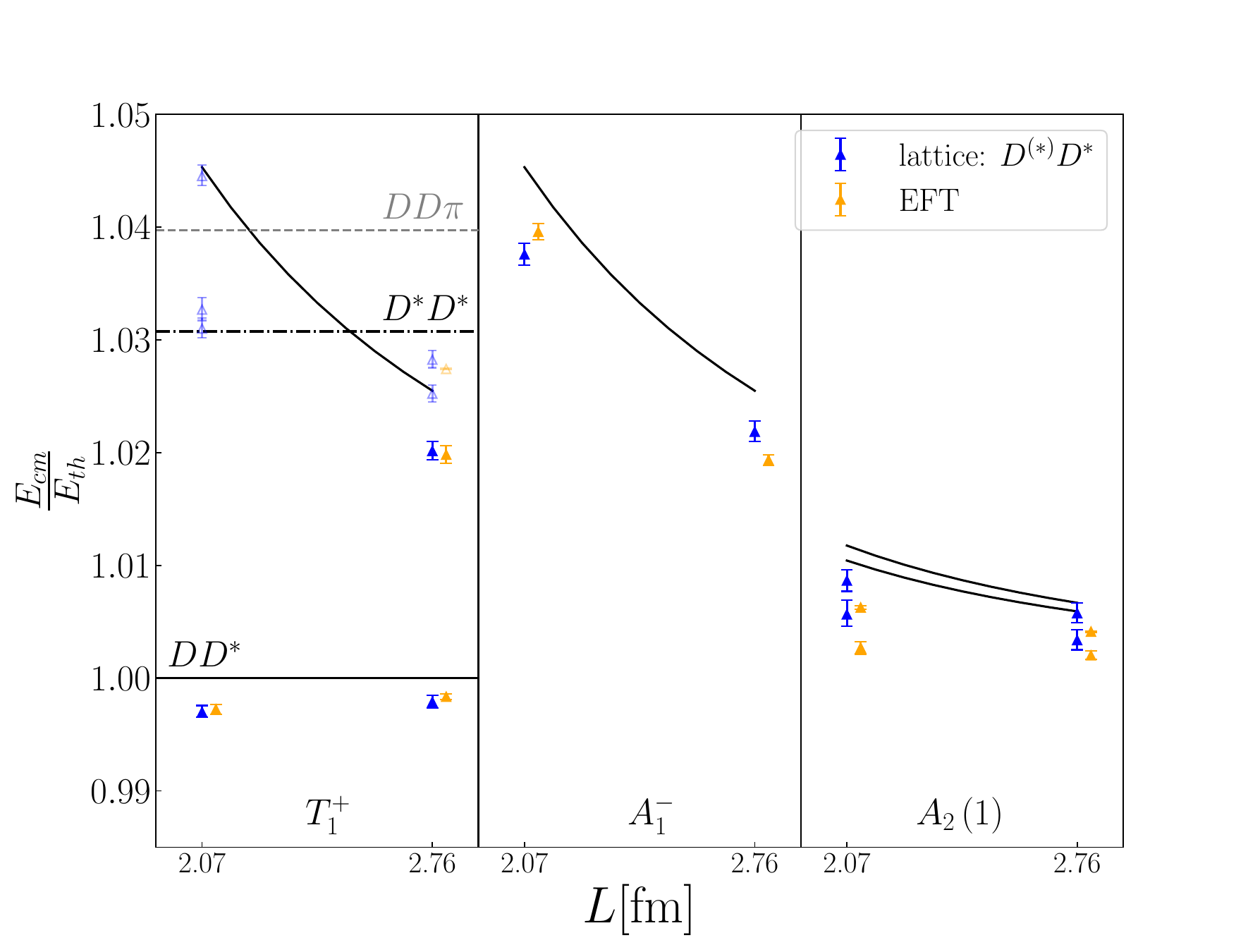}
        \caption{$\Lambda = 0.65~\mathrm{GeV},~n=10,~\frac{\chi^2}{n_{dof}} = 5.5$.}
    \end{subfigure}
    \begin{subfigure}{0.48\textwidth}
        \includegraphics[width=\textwidth]{MM_eft_lattice_0.65_40.pdf}
        \caption{$\Lambda = 0.65~\mathrm{GeV},~n=40,~\frac{\chi^2}{n_{dof}} = 2.4$.}
    \end{subfigure}
    \caption{ Same as Figure \ref{fig:lecfit-app}, but for lattice data based solely on meson-meson operators. }
    \label{fig:lecfit-app-MM}
\end{figure*}

\section{Example of the plane-wave approach for irrep \texorpdfstring{$T_1^+$}{T1p}\label{app:pw}}

\indent This appendix presents a simple application of the plane-wave formalism to the system consisting of a pseudoscalar meson (P) and a vector meson (V). For simplicity, we illustrate the irreducible representation $T_1^+$ of the 48-element Octahedral group $O_h$ for total momentum zero, $\vec{P} = \vec{0}$. In the case of a sharp cut-off on the cmf momentum in the range $\Lambda=\tfrac{2\pi}{L}-\sqrt{2}\tfrac{2\pi}{L}$, the largest momentum shell of plane-waves that contributes to the total basis is $|\vec p|=\tfrac{2\pi}{L}$. Note that this also applies to the channel $PV=DD^*$ considered in the main body of this text with $\Lambda=0.65~$GeV in the case of a sharp regulator $f_{rep}$ in (\ref{freg}). \\
\indent The relevant plane-wave basis $|P(\vec k)V(-\vec k,\vec \epsilon^{~r})\rangle$ (\ref{eq:cmfpw}) is $(1+6)\times 3=21$-dimensional:  
\small{
\begin{equation}
\begin{pmatrix}
P(\vec 0)V_x(\vec 0)\\
P(\vec e_x)V_x(-\vec e_x)\\
P(\vec e_y)V_x(-\vec e_y)\\
P(\vec e_z)V_x(-\vec e_z)\\
P(-\vec e_x)V_x(\vec e_x)\\
P(-\vec e_y)V_x(\vec e_y)\\
P(-\vec e_z)V_x(\vec e_z)\\
 ~\\
 P(\vec 0)V_y(\vec 0)\\
P(\vec e_x)V_y(-\vec e_x)\\
P(\vec e_y)V_y(-\vec e_y)\\
P(\vec e_z)V_y(-\vec e_z)\\
P(-\vec e_x)V_y(\vec e_x)\\
P(-\vec e_y)V_y(\vec e_y)\\
P(-\vec e_z)V_y(\vec e_z)\\
~\\
P(\vec 0)V_z(\vec 0)\\
P(\vec e_x)V_z(-\vec e_x)\\
P(\vec e_y)V_z(-\vec e_y)\\
P(\vec e_z)V_z(-\vec e_z)\\
P(-\vec e_x)V_z(\vec e_x)\\
P(-\vec e_y)V_z(\vec e_y)\\
P(-\vec e_z)V_z(\vec e_z)\\
\end{pmatrix}.
\label{pwT1}
\end{equation}  
}
The representations of the kinetic energy operator $W_{kin}=p^2/2m_r$ and the contact potential $V_{CT}$ (\ref{V}) in this basis form  $21\times 21$ matrices:
\small{
\begin{widetext}
\begin{align}
\label{WVdefs}
\hspace{2.0cm}&W_{kin}=\begin{pmatrix}w_{kin} & 0 & 0\\ 0& w_{kin} & 0\\0& 0 & w_{kin}\end{pmatrix},\qquad V_{CT}=\begin{pmatrix}v_{CT} & 0 & 0\\ 0& v_{CT} & 0\\0& 0 & v_{CT}\end{pmatrix},\nonumber\\ \nonumber \\
&\hspace{-0.2cm}w_{kin}=\frac{1}{2m_r}
\begin{pmatrix}0 & 0 & 0&  0& 0&0& 0\\0 & (\tfrac{2\pi}{L})^2 & 0&  0& 0&0& 0\\0 & 0 & (\tfrac{2\pi}{L})^2&  0& 0&0& 0\\0 & 0 & 0&  (\tfrac{2\pi}{L})^2& 0&0& 0\\0 & 0 & 0&  0& (\tfrac{2\pi}{L})^2&0& 0\\0 & 0 & 0&  0& 0&(\tfrac{2\pi}{L})^2& 0\\0 & 0 & 0&  0& 0&0& (\tfrac{2\pi}{L})^2\end{pmatrix},\nonumber\\ \nonumber \\
&\hspace{-2.5cm}v_{CT}=\begin{pmatrix}
c_0 &c_0+c_2  &c_0+c_2 &  c_0+c_2& c_0+c_2& c_0+c_2&c_0+c_2\\
c_0+c_2 &c_0+2c_2  &c_0+2c_2 &  c_0+2c_2& c_0+2c_2& c_0+2c_2&c_0+2c_2\\
c_0+c_2 &c_0+2c_2  &c_0+2c_2 &  c_0+2c_2& c_0+2c_2& c_0+2c_2&c_0+2c_2\\
c_0+c_2 &c_0+2c_2  &c_0+2c_2 &  c_0+2c_2& c_0+2c_2& c_0+2c_2&c_0+2c_2\\
c_0+c_2 &c_0+2c_2  &c_0+2c_2 &  c_0+2c_2& c_0+2c_2& c_0+2c_2&c_0+2c_2\\
c_0+c_2 &c_0+2c_2  &c_0+2c_2 &  c_0+2c_2& c_0+2c_2& c_0+2c_2&c_0+2c_2\\
c_0+c_2 &c_0+2c_2  &c_0+2c_2 &  c_0+2c_2& c_0+2c_2& c_0+2c_2&c_0+2c_2\\
\end{pmatrix},\quad c_0\equiv c_0^s, ~c_2\equiv c_2^s (\tfrac{2\pi}{L})^2 .  
\end{align}
\end{widetext}
} Note that the $p$-wave contribution in \eqref{WVdefs}, proportional to the low-energy constant $c_2^p$, is omitted since its projection to the considered $T_1^+$ irrep vanishes. \\
\indent General relations required to project the Hamiltonian $H =W_{kin} + V_{CT}$ to arbitrary irreducible representations $\Gamma$ are presented in eqs. (3.5) - (3.10) of \cite{Meng:2021uhz}. To this end, we utilize the matrix $U^{T_{1z}^+}$ that contains the maximally linearly independent set of orthonormal vectors which transform according to the $T_1^+$ irrep. Due to rotational invariance, it suffices to consider only a single row of the irrep, e.g. the $z$-component. The full projection of the 21-dimensional plane-wave basis (\ref{pwT1}) to the basis that transforms accordingly is then encoded by the $3\times 21$ unitary matrix 
\begin{align}
&U^{T_{1z}^+}=\begin{pmatrix} u_x& u_y & u_z\end{pmatrix},\quad u_z=\begin{pmatrix} 
1 & 0 & 0&  0& 0&0& 0\\  0 & 0 & 0&  \tfrac{1}{\sqrt{2}}& 0&0& \tfrac{1}{\sqrt{2}}\\0 & \tfrac{1}{2} &\tfrac{1}{2} &  0&\tfrac{1}{2} &\tfrac{1}{2}& 0\end{pmatrix},\nonumber\\ \nonumber \\
\quad  &u_{x,y}=\begin{pmatrix} 
0 & 0 & 0&  0& 0&0& 0\\  0 & 0 & 0&  0& 0&0& 0\\0 & 0 &0 &  0&0&0& 0\end{pmatrix}.
\end{align}

The Hamiltonian, projected to $T_{1z}^+$, is defined as $H^{T_1^+}=U^{T_{1z}^+} H(U^{T_{1z}^+})^\dagger$ and its eigenvalues are fitted to the lattice energies according to (\ref{eq:hameqirr}). In the absence of one-pion exchange and with definitions in (\ref{WVdefs}), this Hamiltonian then evaluates to 
\begin{align}
&H^{T_1^+}=U^{T_{1z}^+}(W_{kin}+V_{CT}) (U^{T_{1z}^+})^\dagger =\\ \nonumber \\
&=\begin{pmatrix} 
c_0 & \sqrt{2}(c_0+c_2) & 2 (c_0+c_2) \\ \sqrt{2}(c_0+c_2) & 2c_0+4c_2+(\tfrac{2\pi}{L})^2 & 2 \sqrt{2}(c_0+2c_2) \\ 
2(c_0+c_2) & 2 \sqrt{2}(c_0+2c_2) & 4c_0+8c_2+ (\tfrac{2\pi}{L})^2
\end{pmatrix}.\nonumber
\end{align}

\bibliographystyle{utphys-noitalics}
\bibliography{references_MM_4q}

\end{document}